\documentclass[aps,prd,twocolumn,groupedaddress,nofootinbib,superscriptaddress,notitlepage]{revtex4-1}
\usepackage[normalem]{ulem}
\usepackage{verbatim}
\usepackage{graphicx}
\usepackage{dcolumn}
\usepackage{bm}
\usepackage{xcolor}
\usepackage{url}
\usepackage{amsmath}
\usepackage{morefloats}
\usepackage{times}
\usepackage[varg]{txfonts}
\usepackage{multirow}
\usepackage{lineno}
\usepackage{hyperref}
\usepackage{notoccite}
\usepackage{acronym}
\usepackage{xspace}
\usepackage{amssymb}
\usepackage{float}
\usepackage{latexsym}
\usepackage[flushleft]{threeparttable}
\usepackage{natbib}
\usepackage{subfigure}
\usepackage{comment}
\usepackage{booktabs}
\usepackage{placeins}
\usepackage[utf8]{inputenc}

\DeclareGraphicsExtensions{.pdf,.png,.jpg,.jpeg,.eps}
\graphicspath{{./Figures/}}

\def\code#1{\texttt{#1}}

\begin{document}

\title{Cosmological Inference using Gravitational Wave Standard Sirens: A Mock Data Analysis}

\author{Rachel~Gray}
\email{rachel.gray@ligo.org}
\affiliation{SUPA, University of Glasgow, Glasgow G12 8QQ, United Kingdom}
\author{Ignacio~Maga\~{n}a Hernandez}
\email{ignacio.magana@ligo.org}
\affiliation{University of Wisconsin-Milwaukee, Milwaukee, Wisconsin 53201, USA}
\author{Hong~Qi}
\email{hong.qi@ligo.org}
\affiliation{Cardiff University, Cardiff CF24 3AA, United Kingdom}
\author{Ankan~Sur}
\email{ankan.sur@ligo.org}
\affiliation{Nikhef, Science Park 105, 1098 XG Amsterdam, Netherlands}
\affiliation{Nicolaus Copernicus Astronomical Center, Polish Academy of Sciences, 00-716, Warsaw, Poland}
\author{Patrick~R.~Brady}
\affiliation{University of Wisconsin-Milwaukee, Milwaukee, WI 53201, USA}
\author{Hsin-Yu~Chen}
\affiliation{Black Hole Initiative, Harvard University, Cambridge, Massachusetts 02138, USA}
\author{Will~M.~Farr}
\affiliation{Department of Physics and Astronomy, Stony Brook University, Stony Brook, New York 11794, USA}
\affiliation{Center for Computational Astronomy, Flatiron Institute, New York 10010, USA}
\author{Maya~Fishbach}
\affiliation{University of Chicago, Chicago, Illinois 60637, USA}
\author{Jonathan~R.~Gair}
\affiliation{School of Mathematics, University of Edinburgh, Edinburgh EH9 3FD, United Kingdom}
\affiliation{Max Planck Institute for Gravitational Physics (Albert Einstein Institute), Potsdam-Golm, 14476, Germany}
\author{Archisman~Ghosh}
\affiliation{Nikhef, Science Park 105, 1098 XG Amsterdam, Netherlands}
\affiliation{Delta Institute for Theoretical Physics, Science Park 904, 1090 GL Amsterdam, Netherlands}
\affiliation{Lorentz Institute, Leiden University, PO Box 9506, Leiden 2300 RA, Netherlands}
\affiliation{GRAPPA, University of Amsterdam, Science Park 904, 1098 XH Amsterdam, Netherlands}
\author{Daniel~E.~Holz}
\affiliation{University of Chicago, Chicago, IL 60637, USA}
\author{Simone~Mastrogiovanni}
\affiliation{Laboratoire Astroparticule et Cosmologie, CNRS, Universit\'e de Paris, 75013 Paris, France}
\author{Christopher~Messenger}
\affiliation{SUPA, University of Glasgow, Glasgow G12 8QQ, United Kingdom}
\author{Danièle~A.~Steer}
\affiliation{Laboratoire Astroparticule et Cosmologie, CNRS, Universit\'e de Paris, 75013 Paris, France}
\author{John~Veitch}
\affiliation{SUPA, University of Glasgow, Glasgow G12 8QQ, United Kingdom}

\date{\today}
\begin{abstract}
The observation of binary neutron star merger GW170817, along with its optical 
counterpart, provided the first constraint on the Hubble constant $H_0$  
using gravitational wave standard sirens.
When no counterpart is identified, a galaxy catalog can be used 
to provide the necessary redshift information. However, the true host might not 
be 
contained in a catalog which is not complete out to the limit of 
gravitational-wave detectability.
These electromagnetic and gravitational-wave
selection effects must be accounted for. We describe and implement a method to 
estimate $H_0$ using both
the counterpart and the galaxy catalog standard siren methods. We perform a 
series 
of mock data analyses using binary neutron star mergers to confirm our 
ability to recover an unbiased estimate of $H_0$. Our simulations 
used a simplified universe with no redshift uncertainties or galaxy clustering, 
but with different magnitude-limited catalogs and assumed host galaxy 
properties, to test our treatment of both selection effects.
We explore how the incompleteness of 
catalogs affects the final measurement of $H_0$, as well as the 
effect of weighting each galaxy's likelihood of being a host by its 
luminosity.
In our most realistic simulation, 
where the simulated catalog is about three times denser 
than the density of galaxies in the local universe,
we find that a 4.4\% measurement precision can 
be reached using galaxy catalogs with 50\% completeness and $\sim 250$ binary neutron 
star detections with sensitivity similar to that of Advanced LIGO's second 
observing run.

\end{abstract}


\maketitle
\acrodef{H0}[$H_0$]{the Hubble constant}
\acrodef{LVC}[LVC]{LIGO and Virgo Collaborations}
\acrodef{GW}[GW]{gravitational waves}
\acrodef{EM}[EM]{electromagnetic}
\acrodef{CMB}[CMB]{cosmic microwave background}
\acrodef{SN}[SN]{supernovae}
\acrodef{MDC}[MDA]{mock data analysis}
\acrodefplural{MDC}[MDAs]{mock data analyses}
\acrodef{BNS}[BNS]{binary neutron star}
\acrodef{BBH}[BBH]{binary black hole}
\acrodef{LCDM}[$\Lambda$CDM]{$\Lambda$-cold-dark-matter}
\acrodef{SNR}[SNR]{signal-to-noise ratio}
\acrodef{O2}[O2]{observing run 2}
\acrodef{O3}[O3]{observing run 3}
\acrodef{F2Y}[F2Y]{First Two Years}
\acrodef{GRB}[GRB]{gamma-ray burst}
\acrodef{SH0ES}[SH0ES]{Supernovae, $H_0$, for the Equation of State of Dark energy}
\acrodef{O1}[O1]{the first observing run}
\acrodef{O2}[O2]{the second observing run}
\acrodef{O3}[O3]{the third observing run}
\acrodef{NSBH}[NSBH]{neutron star - black hole pair}
\acrodef{CBC}[CBC]{compact binary coalescence}

\def\lalinference{\textsc{LALInference}\ }
\def\bomega{{\bf \Omega}}
\def\btheta{{\mathbf{\theta}}}
\def\mce{\mathcal E}
\def\mch{\mathcal H}
\def\mci{\mathcal I}
\def\mcn{\mathcal N}
\def\ciota{{\cos\iota}}
\def\sdelta{{\sin\!\delta}}
\def\pint{p_{\text{int}}}
\def\edet{{\mce_\text{det}}}
\def\neff{{\mcn_\text{eff}}}
\def\pdet{p_{\text{det}}}
\def\rhoth{{\rho_\text{th}}}
\def\Mpc{\text{Mpc}}
\def\Mz{\ensuremath{\mathcal{M}_z}}

\newcommand{\kmsMpc}{\ensuremath{\mbox{km s}^{-1} \,\mbox{Mpc}^{-1}}}
\newcommand{\sqdeg}{\ensuremath{\text{deg}^2}\xspace}
\newcommand{\Hubblemdccounter}{\ensuremath{69.08^{+0.79}_{-0.80}}}
\newcommand{\HubblemdccounterMAP}{\ensuremath{69.08}}
\newcommand{\Hubblemdccountererror}{\ensuremath{1.13}}
\newcommand{\Hubblemdc}{\ensuremath{68.91^{+1.36}_{-1.22}}}
\newcommand{\HubblemdcMAP}{\ensuremath{68.91}}
\newcommand{\Hubblemdcerror}{\ensuremath{1.84}}
\newcommand{\Hubblemdca}{\ensuremath{70.14^{+2.29}_{-2.18}}}
\newcommand{\HubblemdcaMAP}{\ensuremath{70.14}}
\newcommand{\Hubblemdcaerror}{\ensuremath{3.20}}
\newcommand{\Hubblemdcb}{\ensuremath{70.14^{+1.80}_{-1.67}}}
\newcommand{\HubblemdcbMAP}{\ensuremath{70.14}}
\newcommand{\Hubblemdcberror}{\ensuremath{2.48}}
\newcommand{\Hubblemdcc}{\ensuremath{69.97^{+1.59}_{-1.50}}}
\newcommand{\HubblemdccMAP}{\ensuremath{69.97}}
\newcommand{\Hubblemdccerror}{\ensuremath{2.21}}
\newcommand{\Hubblemdcweights}{\ensuremath{70.83^{+3.55}_{-2.72}}}
\newcommand{\HubblemdcweightsMAP}{\ensuremath{70.83}}
\newcommand{\Hubblemdcweightserror}{\ensuremath{4.48}}
\newcommand{\Hubblemdcnoweights}{\ensuremath{69.50^{+4.20}_{-3.24}}}
\newcommand{\HubblemdcnoweightsMAP}{\ensuremath{69.50}}
\newcommand{\Hubblemdcnoweightserror}{\ensuremath{5.31}}
\newcommand{\priorrange}{\ensuremath{[20,200] \ \kmsMpc}}
\newcommand\note[1]{\textcolor{red}{#1}}

\section{Introduction\label{sec:introduction}}
%
The idea that \ac{GW} detections can be used for the inference of cosmological
parameters, such as \ac{H0}, was first proposed over three decades
ago by Bernard Schutz~\cite{1986Natur.323..310S}. The key to this process is that \ac{GW}
signals from \acp{CBC} act as standard sirens, in the sense that they provide a
self-calibrated luminosity distance to the source. This can be obtained
directly from the \ac{GW} signal, and is therefore entirely independent of the
cosmic distance ladder~\cite{Holz:2005df,Dalal:2006qt,Nissanke:2009kt,Nissanke:2013fka,Chen:2017rfc,Feeney:2018mkj,DiValentino:2018jbh,Mortlock:2018azx,Farr:2019twy}. With the addition of redshift information for each
source we then have the required input for cosmological
inference. 

%
At the time of writing, the current percent level state-of-the-art \ac{EM} measurements of
\ac{H0} are in tension with each other. The Planck experiment uses measurements of \ac{CMB}
anisotropies and provides a value of {$H_0 = 67.4 \pm 0.5$
\kmsMpc}~\cite{Aghanim:2018eyx}. The \ac{SH0ES} experiment measures distances to
Type Ia supernovae standard candles making use of the cosmic distance ladder,
and gives {$H_0 = 74.03 \pm 1.42$ \kmsMpc}~\cite{Riess:2019cxk}. These two
independent measurements of \ac{H0} are in tension at the level of {$\sim
4.4$-$\sigma$}~\cite{Riess:2019cxk}.
While the early-universe Planck measurements are also favored by
measurements using supernovae calibrated with Baryon Acoustic
Oscillations~\cite{Macaulay:2018fxi}, and the SH0ES results agree with
local gravitational lensing
measurements by the H0LiCOW Collaboration~\cite{Birrer:2018vtm},
calibration of supernovae using the Tip of the Red Giant
Branch yields $H_0$ midway between the two~\cite{Freedman:2019jwv}.

%
This indicates the possibility that at least one of these measurements is
subject to unknown systematics, or it could be an indication of new
physics causing the discrepancy between the local measurements and
the non-local (early universe) CMB based measurement. This makes a \ac{GW} standard siren
measurement of \ac{H0} particularly interesting, as this will provide an alternative local constraint on \ac{H0}.
In this manner, the use of \acp{GW} as standard
sirens may allow us to arbitrate the current situation, indicating either a
bias in the current measurements, or pointing towards new physics.

%
The detection of the \ac{BNS} event GW170817~\cite{GW170817:discovery},
together with its optical counterpart~\citep{GW170817:MMA,GW170817:DES} led to
the first standard siren measurement of \ac{H0}~\cite{GW170817:H0}. The
counterpart associated with GW170817 allowed for the identification of its host
galaxy, NGC4993, and hence a direct measurement of its redshift, which in turn
resulted in the inferred value \ac{H0}=$70^{+12}_{-8}$ \kmsMpc.
Future counterpart standard siren measurements are expected to constrain $H_0$
to the percent level~\cite{Dalal:2006qt,Nissanke:2009kt,Nissanke:2013fka,Chen:2017rfc,Feeney:2018mkj}.

%
Central to the aims of this paper is the case where an \ac{EM} counterpart is
not observed, and how \ac{H0} inference can still be performed. In particular,
the method proposed by Schutz in 1986~\cite{1986Natur.323..310S,DelPozzo:2012}
allows the use of galaxy catalogs to provide redshift information for potential
host galaxies within the event's \ac{GW} sky-localization. The idea is that, by
marginalizing over the possible discrete values of redshift for each \ac{GW}
detection we account for uncertainty as to which galaxy is the true host. By
combining the information from many \ac{GW} events, the contributions from the
true host galaxies will grow since they will all share the same true \ac{H0}.
Contributions from the others will statistically average out, leading to a
constraint on \ac{H0} and possibly other cosmological parameters.

%
Over the course of \ac{O1} and \ac{O2} a total of 11 \ac{GW} events were
detected by the advanced LIGO and Virgo detectors: 10 are \ac{BBH} events and
one is the above-mentioned \ac{BNS} event GW170817~\cite{O2:catalog}. The ``galaxy catalog'' method has
been independently applied to both the \ac{BNS} event GW170817 (without
assuming NGC4993 is the host)~\cite{2018arXiv180705667F}, and the \ac{BBH}
event GW170814~\cite{DES} resulting in posterior probability distributions on
\ac{H0} where the posterior from GW170814 was broader than (but consistent with)
that obtained from GW170817. The difference in the widths of the \ac{H0}
constraints is an expected result due to the larger localization volume
associated with GW170814, and the high number of galaxies it contained. Using
the detections from \ac{O1} and \ac{O2}, multiple \ac{GW}
events have been combined to give the latest standard siren measurement of
\ac{H0}~\cite{O2H0paper} using the methodology presented in this paper.

%
Predictions suggest that it will be possible to constrain \ac{H0} to less than
$2\%$ within 5 years of the start of \ac{O3} and to $1\%$ within a decade,
though this is dependent on the number of events observed with \ac{EM}
counterparts~\cite{Chen:2017rfc}, and this may change as our understanding of
astrophysical rates improves, and would require the detector amplitude calibration
error to be measured to better than this precision.
Simulations in~\cite{Chen:2017rfc} and~\cite{2018arXiv180705667F},
which assume complete catalogs based on realistic large-scale structure simulations,
find that for BNSs without counterparts, the convergence is $40\%/\sqrt{N}$.
The convergence found there for BBHs is much slower, as \acp{BBH} are typically
detected at greater distances with larger localization volumes.

The prospects of identifying a transient \ac{EM} counterpart will certainly increase,
and correspondingly, the number of candidate host galaxies in a catalog
will decrease, with improved event sky-localizations as future \ac{GW}
observatories join the detector network~\cite{Aasi:2013wya}.
With the Japanese detector KAGRA~\cite{Aasi:2013wya}
 having joined \ac{O3} in early 2020, and LIGO-India approved for
construction~\cite{LIGOIndia}, the next decade of standard siren cosmology is
set to be very exciting.

%
\ac{O3} began on April 1\textsuperscript{st} 2019 and consists of 11 months' worth of data.
 The sensitivities of the LIGO and Virgo detectors have improved since
\ac{O2}, leading to an increased detection rate of \ac{GW}
candidates\footnote{In the first half of \ac{O3} the detectors averaged
the detection of one \ac{GW} candidate per week. If all of these candidates are
ultimately identified as real \ac{GW} events, then \ac{O3} within its first two
months will have exceeded
the total number of detections of \ac{O1} and \ac{O2}.}~\cite{O2:Rates}. This is the first observing run for which there will
be 3 detectors operating for the entirety of the run. Having more detectors improves
the duty-cycle of the network, {\em i.e.}~the fraction of run time
for which one or more detectors in the network is online, and
also increases the rate of three-detector detections, which will
likely be better localized on the sky than the two-detector ones.
This is important, both in terms of performing \ac{EM} follow-up for
\ac{EM} counterparts practically~\cite{2019ApJ...875..161A}, and for reducing the
number of possible host galaxies for events in the case where a counterpart is
not observed.

%
This paper presents the Bayesian framework behind the \code{gwcosmo} code, a
product of the \ac{LVC} which was used to measure \ac{H0} using detected
\ac{GW} events from \ac{O1} and \ac{O2}~\cite{O2H0paper}. 
 The method detailed in this paper is also expected to be implemented in future LIGO/Virgo/KAGRA standard siren measurements.
We present results from a series of
\acp{MDC} which were designed specifically to test this method's robustness
against some of the most common pitfalls, in particular, \ac{GW} selection
effects which affect all $H_0$ measurements, and \ac{EM} selection effects,
which are relevant in the context of galaxy catalogs. This method builds upon the
Bayesian framework first presented in~\cite{DelPozzo:2012} which has
subsequently been extended, modified and independently derived by multiple
authors~\cite{Nissanke:2013fka,Chen:2017rfc,Nair:2018ign,2018arXiv180705667F,DES}. The framework here is broadly equivalent to that in~\cite{Chen:2017rfc,2018arXiv180705667F}, however the mathematics and implementation differ, most notably in the treatment of \ac{EM} selection effects. With specific care regarding
selection effects we outline methods for constraining \ac{H0} using both the
``galaxy catalog'' and ``\ac{EM} counterpart'' approaches. 

%
This paper is the first to explicitly test the robustness of a coded implementation of this methodology through use of galaxy catalogs which are incomplete and do not contain all of the \ac{GW} host galaxies.  Additionally, the \ac{GW} data used in these \acp{MDC} were produced using an end-to-end simulation, including searching for ``injected'' signals in real detector data followed by a full parameter estimation to obtain the \ac{GW} posterior samples~\cite{2014ApJ...795..105S,Berry:2014jja}, making this the most realistic set of simulated \ac{GW} data to be used to explore \ac{GW} cosmology to date.  The analyses start with the most simplistic scenario, and increase in complexity with each iteration in order to ensure that the \code{gwcosmo} code is able to pass each level satisfactorily before moving onto the next. 

%
This paper is structured as follows. Section~\ref{sec:methodology} presents
the Bayesian framework used to estimate the posterior on \ac{H0}.
Section~\ref{sec:mdc} discusses the design and preparation of the \acp{MDC}.
In Section~\ref{sec:results} we present our results. We conclude in
Section~\ref{sec:discussion} giving a detailed discussion of results and
providing guidance for future work. 
Some of the details of the Bayesian method have been set aside to be discussed in an Appendix.

\section{Methodology\label{sec:methodology}}
The late-time cosmological expansion in a Friedmann-Lema\^itre-Robertson-Walker universe is characterized by the Hubble-Lema\^itre parameter as a function of the redshift $z$,
 \begin{equation}\label{Eq:hubblez}
    H(z) = H_0 \sqrt{\Omega_\text{m}(1+z)^3 +\Omega_\text{k}(1+z)^2 + \Omega_\Lambda}\,,
\end{equation}
where $H_0$ is the Hubble constant, the rate of expansion in the current epoch, and $\Omega_\text{m}$ and $\Omega_\Lambda$ are the fractional matter density (including baryonic and cold dark matter) and fractional dark energy density (assumed to be due to a cosmological constant) respectively; $\Omega_\text{k}$ is the fractional curvature energy density  which is identically zero for a ``flat'' universe consistent with observations. Additionally, we have the constraint $\Omega_\text{m} + \Omega_\text{k} + \Omega_\Lambda = 1$ for all the components contributing to the energy density of universe at the present epoch.

The expansion history of the universe maps to a ``redshift-distance relation'' associating the redshift $z$ of observable sources to their luminosity distance $d_L(z)$ (see {\em e.g.}~\cite{Hogg:1999ad}) as,
\begin{eqnarray}
d_L(z)&=&\frac{c \, (1+z)}{H_0} \int_0^{z} \, \frac{H_0}{H(z')} \, dz'\,,
\label{eqn:dLz}
\end{eqnarray}
for a flat universe.
From the relation between observed $z$ and $d_L$ to sources (\ac{EM} sources such as variable stars or supernovae, or \ac{GW} sources), one can measure the cosmological parameters appearing in $H(z)$. With knowledge of the other cosmological parameters $\{ \Omega_\text{m}, \Omega_\text{k}, \Omega_\Lambda\}$ coming from independent observations, the redshift-distance relation can be used to measure \ac{H0}.
We would like to note that with
prior knowledge on the other cosmological parameters coming from \ac{EM} observations, the measurement made with \ac{GW} detections are not strictly independent measurements.

At low redshifts $z\ll 1$, the redshift-distance relation can be approximately described by the linear Hubble relation,
\begin{equation}\label{Eq:Hubble}
    d_L(z) \approx c\,z/H_0\,,
\end{equation}
which contains \ac{H0} but is independent of the other cosmological parameters.
With this approximate linear relation at low redshifts, any measurement of $H_0$ with GWs is independent of the values of the other cosmological parameters.

\subsection{Standard Sirens\label{sec:sirens}}
The amplitude of the observed strain is inversely proportional to the luminosity distance to the \ac{GW} source. For compact binary sources in quasi-circular orbits, the two polarizations of the gravitational wave signal can be written to leading order as a function of frequency $f$ as ~\cite{Maggiore}
\begin{align}
\tilde{h}_+(f)&\propto \frac{\Mz^{5/6}}{2d_L}\left(1+\cos^2(\iota)\right)f^{-7/6}\exp\left(i\phi(\Mz,f)\right) \\
\tilde{h}_\times(f)&\propto \frac{\Mz^{5/6}}{d_L}\cos(\iota)f^{-7/6}\exp\left(i\phi(\Mz,f)+i\pi/2\right)
\label{Eq:GWdist}
\end{align}
where $\phi(\mathcal{M}_z,t)$ is the phase of the signal.
The redshifting of the signal is accounted for by using the parameter $\Mz \equiv \mathcal{M}(1+z)$,
the ``redshifted chirp mass,'' to describe the signal as observed in the detector.
Since $\mathcal{M}_z$ appears in both the phase and the amplitude, and in practice is
more strongly constrained by $\phi(\mathcal{M}_z,f)$, the dominant uncertainty on the
signal amplitude results from the uncertainties on luminosity distance, $d_L$ and inclination angle $\iota$. Each detector sees a linear combination of the two polarizations,
$\tilde{h}(t)=F_+\tilde{h}_+ + F_\times \tilde{h}_\times$, where $F_{+,\times}$ are the antenna response functions of the detector, which vary over the sky position and polarization
angle of the source. Given multiple detectors at distant sites it is possible to
simultaneously infer the parameters of the source, and therefore find a direct estimate
of its luminosity distance\,\cite{LALInference}. This makes compact binaries self-calibrated luminosity distance indicators or ``standard sirens'' unlike \ac{EM} distance indicators which need to undergo calibration via multiple rungs of the cosmic distance ladder. The redshift of the \ac{GW} source, also required for cosmological inference, remains degenerate with the source's mass, contained within $\mathcal{M}_z$, and needs to be estimated in alternate ways. The precision of the $d_L$ estimate is limited because of correlations with other
parameters, particularly the inclination angle $\iota$~\cite{Nissanke:2013fka}.
In this work we simulate these effects as part of our end-to-end analysis, described
in Sec.~\ref{sec:mdc}.

\subsection{Galaxy Information\label{sec:galaxies}}
There are multiple ways in which \ac{EM} observations can provide complementary redshift\footnote{There are ways of obtaining the redshift independent of \ac{EM} observations, by using known population properties such as the mass distribution \cite{massdist,Farr:2019twy}, or the neutron star equation-of-state \cite{eos}.} information. A \ac{BNS} event may be detected in coincidence with an \ac{EM} counterpart, which can be associated with the host galaxy to provide a direct measurement of the redshift of the source. More generically, a \ac{GW} event may not have a detected \ac{EM} counterpart, in which case one  needs to fall back on the method outlined by Schutz~\cite{1986Natur.323..310S} and use potential host galaxies within the event's sky localization region for the redshift information for the source. Two possibilities come up: (i) to use available galaxy catalogs, or (ii) to conduct dedicated \ac{EM} follow-up on the event's sky region, mapping the galaxies within that area to as great a depth as possible to maximize the redshift information available.

When using galaxy catalogs to provide the prior redshift information, the possibility that the host galaxy lies beyond the reach of the catalog must be taken into account.  \ac{EM} telescopes are flux limited, which means that galaxy catalogs are inherently biased towards containing objects which are brighter and/or nearer-by (although there may be other selection effects due to galaxy color or size, depending on the catalog).  
These \ac{EM} selection effects must be accounted for. Carrying out dedicated \ac{EM} follow-up will, to some degree, mitigate this issue, as it will allow for far deeper coverage over a small section of the sky.  For nearby events, the possibility that the host galaxy lies above the telescope's upper threshold may be negligible.  However, the time and resources required for dedicated \ac{EM} follow-up means that the default approach for \ac{GW} events observed without counterparts will be to use pre-existing catalogs.

In either case, the uncertainty associated with each galaxy's redshift must be taken into account, including the redshift error due to the galaxy's peculiar velocity, $v_p$, and, in cases where the redshift is estimated photometrically, a much larger uncertainty due to the photometric algorithm.
Peculiar velocities are significant for nearby galaxies. 
The effect of the peculiar velocity on the measurement of \ac{H0} may be small if there are a large number of potential host galaxies in the \ac{GW} event's sky-localization, but for a small number of galaxies, and for the counterpart case, this effect is particularly noticeable.
For GW170817 at a nearby distance of about $40\,\text{Mpc}$, the peculiar velocity contribution was large as $10\%$ of the total observed redshift \cite{GW170817:H0}, and different procedures of reconstructing the peculiar velocity field led to residual uncertainties on the redshift of between $2\%$ and $8\%$ \cite{GW170817:H0,Springob:2014qja,Carrick:2015xza,Hjorth:2017yza,Guidorzi:2017ogy}. The impact on $H_0$ measurement of peculiar velocities and their reconstruction is of topical interest, and has been the subject of several recent studies including \cite{Howlett:2019mdh,Mukherjee:2019qmm,Nicolaou:2019cip}.
Photometric redshifts on the other hand are important slightly farther away
due to lack of spectroscopic data in galaxy catalogs. The ``photo-$z$'' are estimated using fitting and machine learning algorithms \cite{DeVicente:2015kyp,Sadeh:2015lsa}, which often have large $\mathcal{O}(1)$ fractional uncertainties associated with them. While various caveats and subtleties for a realistic measurement have been outlined in \cite{O2H0paper}, the impact of photo-$z$ uncertainties on $H_0$ measurement is not precisely quantified in literature yet. Our present mock data analyses ignore these crucial redshift uncertainties altogether, and the impact of their magnitudes, profiles, and other systematic artefacts are left aside for possible future study.

\subsection{Bayesian Framework \label{Sec: Overview}}
{\renewcommand{\arraystretch}{1.5}
\begin{table}[ht]
\centering
\begin{ruledtabular}
\begin{tabular}{p{1.5cm}p{6.5cm}}
Parameter & Definition\\ \hline \hline
$H_0$ & The Hubble constant. \\ \hline
$N_{\text{det}}$ & The number of events detected during the observation period. \\ \hline
$x_{\text{GW}}$ & The \ac{GW} data associated with some \ac{GW} source, $s$.  \\ \hline
$D_{\text{GW}}$ & Denotes that a \ac{GW} signal was detected, {\em i.e.}~that $x_{\text{GW}}$ passed some detection statistic threshold $\rho_\text{th}$. \\ \hline
$g$ & Denotes that a galaxy is ($G$), or is not ($\bar{G}$), contained within the galaxy catalog. \\ \hline
$x_{\text{EM}}$ & The \ac{EM} data associated with some \ac{EM} counterpart. \\ \hline
$D_{\text{EM}}$ & Denotes that an \ac{EM} counterpart was detected, {\em i.e.}~that $x_{\text{EM}}$ passed some threshold. \\ 
\end{tabular}
\end{ruledtabular}
\caption{\label{tab:params} A summary of the parameters present in the methodology.}
\end{table}
}
This section presents an overview of the Bayesian framework of the \code{gwcosmo} methodology.  Parameters which appear explicitly in this overview are defined in Table \ref{tab:params},
while Table \ref{tab:params_ext} in Appendix \ref{Sec: Components} provides an extended list of parameter definitions, alongside a network diagram which demonstrates the conditional dependence of these parameters (see Fig.~\ref{fig:network}).

The posterior probability on $H_0$ from $N_{\text{det}}$ \ac{GW} events is computed as follows:
\begin{equation}\label{Eq:posteriormain}
p(H_0|\{x_{\text{GW}}\},\{D_{\text{GW}}\})\propto p(H_0)p(N_{\text{det}}|H_0)\prod_i^{N_{\text{det}}} p({x_{\text{GW}}}_i|{D_{\text{GW}}}_i,H_0)
\end{equation}
where $\{x_{\text{GW}}\}$ is the set of \ac{GW} data, $D_{\text{GW}}$ indicates
that the event was detected as a \ac{GW} and $p(H_0)$ is the prior on $H_0$.
For a given $H_0$, the term $p(N_{\text{det}}|H_0)$ is the probability
of detecting $N_{\text{det}}$ events. It depends on the intrinsic astrophysical
rate of events in the source frame, $R=\frac{\partial{N}}{\partial V \partial
T}$. The total number of expected events is given by $N_{\text{det}}=R\,
\langle VT \rangle$, where $\langle VT \rangle$ is the average of the
surveyed comoving volume multiplied by the observation time. By choosing a scale-free
 prior on rate, $p(R) \propto 1/R$, the dependence on $H_0$
drops out~\cite{Fishbach:2018edt}. For simplicity this approximation is made
throughout the analysis and therefore $p(N_{\text{det}}|H_0)$ is absent from
further expressions.

The remaining term factorizes into likelihoods for each detected event. Using Bayes' theorem we can write it as,
\begin{equation}
\label{Eq.xD}
\begin{aligned}
p(x_{\text{GW}}|D_{\text{GW}},H_0) &= \dfrac{p(D_{\text{GW}}|x_{\text{GW}},H_0)p(x_{\text{GW}}|H_0)}{p(D_{\text{GW}}|H_0)}
\\ &= \dfrac{p(x_{\text{GW}}|H_0)}{p(D_{\text{GW}}|H_0)},
\end{aligned} 
\end{equation}
where we set $p(D_{\text{GW}}|x_{\text{GW}},H_0)=1$, since the analysis is only carried out when the \ac{SNR}, $\rho$, associated with $x_{\text{GW}}$ passes some detection statistic threshold $\rho_\text{th}$ -- it is a prerequisite that the event has been detected. Calculating $p(D_{\text{GW}}|H_0)$ requires integrating over all possible realizations of \ac{GW} events, with a lower integration limit of $\rho_\text{th}$:
\begin{equation} \label{Eq:D_H0}
\begin{aligned}
p(D_{\text{GW}}|H_0) = \int_{\rho>\rho_{\text{th}}}^\infty p(x_{\text{GW}}|H_0)dx_{\text{GW}}.
\end{aligned}
\end{equation}
For explicit details on the calculation of $p(D_{\text{GW}}|H_0)$ see Appendix \ref{Ap:GWselection}. 
The term $p(D_{\text{GW}}|H_0)$ depends on properties of the \ac{GW} source population ({\em e.g.} the mass distribution), but in this work, for simplicity, it is assumed that the population properties are known exactly.

\subsubsection{The galaxy catalog method \label{subsec:galcat method}}
In the galaxy catalog case, the \ac{EM} information enters the analysis as a prior,  made up of a series of possibly smoothened delta functions\footnote{While uncertainties on the galaxy sky-coordinates can be safely ignored, the error on the redshift can be modeled with a Gaussian or a more complicated distribution.} at the redshift, right ascension (RA) and declination (dec) of the possible source locations.  As we are in the regime where (especially for \acp{BBH}) galaxy catalogs cannot be considered complete out to the distances to which \ac{GW} events are detectable, we have to consider the possibility that the host galaxy is not contained within the galaxy catalog due to being dimmer than the apparent magnitude threshold.
In order to do so, we marginalize the likelihood over the case where the host galaxy is, and is not, in the catalog (denoted by $G$ and $\bar{G}$ respectively):
\begin{equation} \label{Eq:sum G}
\begin{aligned}
p(x_{\text{GW}}|D_{\text{GW}},H_0) &= \sum_{g=G,\bar{G}} p(x_{\text{GW}}|g,D_{\text{GW}},H_0) p(g|D_{\text{GW}}, H_0), \\
&= p(x_{\text{GW}}|G,D_{\text{GW}},H_0) p(G|D_{\text{GW}}, H_0) \\
&\;\;\; + p(x_{\text{GW}}|\bar{G},D_{\text{GW}},H_0) p(\bar{G}|D_{\text{GW}}, H_0)\,.
\end{aligned} 
\end{equation}
While theoretically equivalent to and consistent with the methodology presented in~\cite{Chen:2017rfc,2018arXiv180705667F}, the mathematics and implementation here differ, most notably in the treatment of \ac{EM} selection effects, and our focus on whether the host galaxy is contained within the galaxy catalog or not, rather than calculating a ``completeness fraction'' in order to weight the in-catalog and out-of-catalog likelihood contributions.  This, alongside the modeling of \ac{EM} selection effects using an apparent magnitude threshold, which has not been done before, accounts for the main differences between this derivation and those presented in earlier works. The methodology presented here aligns directly with the implementation of the \code{gwcosmo} code.  We leave the details of this derivation to Appendix~\ref{Sec: Components}.

\subsubsection{The counterpart method}
The method outlined above is for the galaxy catalog case, in which no \ac{EM} counterpart is observed, or expected. We also consider the case where we observe an \ac{EM} counterpart.  The main difference is the inclusion of a likelihood term for the \ac{EM} counterpart data, mirroring that of the \ac{GW} data.

The likelihood in this case, which is the term within the product in Eq.~(\ref{Eq:posteriormain}), is given by:
\begin{equation}\label{Eq:counterpart}
\begin{aligned}
p(x_{\text{GW}},&x_{\text{EM}}|D_{\text{GW}},D_{\text{EM}},H_0) \\&= \dfrac{p(x_{\text{GW}},x_{\text{EM}}|H_0) p(D_{\text{GW}},D_{\text{EM}}|x_{\text{GW}},x_{\text{EM}},H_0)}{p(D_{\text{GW}},D_{\text{EM}}|H_0)},
\\&= \dfrac{p(x_{\text{GW}}|H_0)p(x_{\text{EM}}|H_0)}{p(D_{\text{EM}}|D_{\text{GW}},H_0)p(D_{\text{GW}}|H_0)}.
\end{aligned} 
\end{equation}
where $x_{\text{EM}}$ refers to the \ac{EM} counterpart data and $D_{\text{EM}}$ denotes that the counterpart was detected. In the numerator we have assumed that the \ac{GW} and \ac{EM} data are independent of each other and so the joint \ac{GW}-\ac{EM} likelihood factors out. $p(D_{\text{GW}},D_{\text{EM}}|x_{\text{GW}},x_{\text{EM}},H_0)$ is further factorized as $p(D_{\text{EM}}|D_{\text{GW}},x_{\text{GW}},x_{\text{EM}},H_0) p(D_{\text{GW}}|x_{\text{GW}},x_{\text{EM}},H_0)$.  The first term is equal to 1, as this method is only used when we have observed an \ac{EM} counterpart, meaning that by definition $x_{\text{EM}}$ has passed some threshold for detectability set by \ac{EM} telescopes.  The second term also goes to 1, due to the same threshold argument as in section \ref{Sec: Overview}.

For simplicity, in this paper we make the assumption that the detection of an \ac{EM} counterpart is flux-limited and, as in~\cite{GW170817:H0}, that the detectability of \ac{EM} counterparts extends well beyond the distance to which \acp{BNS} are detectable with \ac{O2}-like LIGO and Virgo sensitivity. Following this, we make the assumption that the term $p(D_{\text{EM}}|D_{\text{GW}},H_0) \approx 1$, and leave a more rigorous analysis of the $H_0$-dependence of this term for a future study.

In an ideal scenario, the observation of an \ac{EM} counterpart will allow for the identification of one of the galaxies in the neighboring region as the host of the \ac{GW} event. In the case where the \ac{EM} counterpart cannot be unambiguously linked to a host galaxy, this uncertainty can also be taken into account.  See Appendix \ref{Ap:counterpart} for more details.

\section{The Mock Data Analyses\label{sec:mdc}}
\begin{figure*}[htb]
\includegraphics[width=0.48\textwidth]{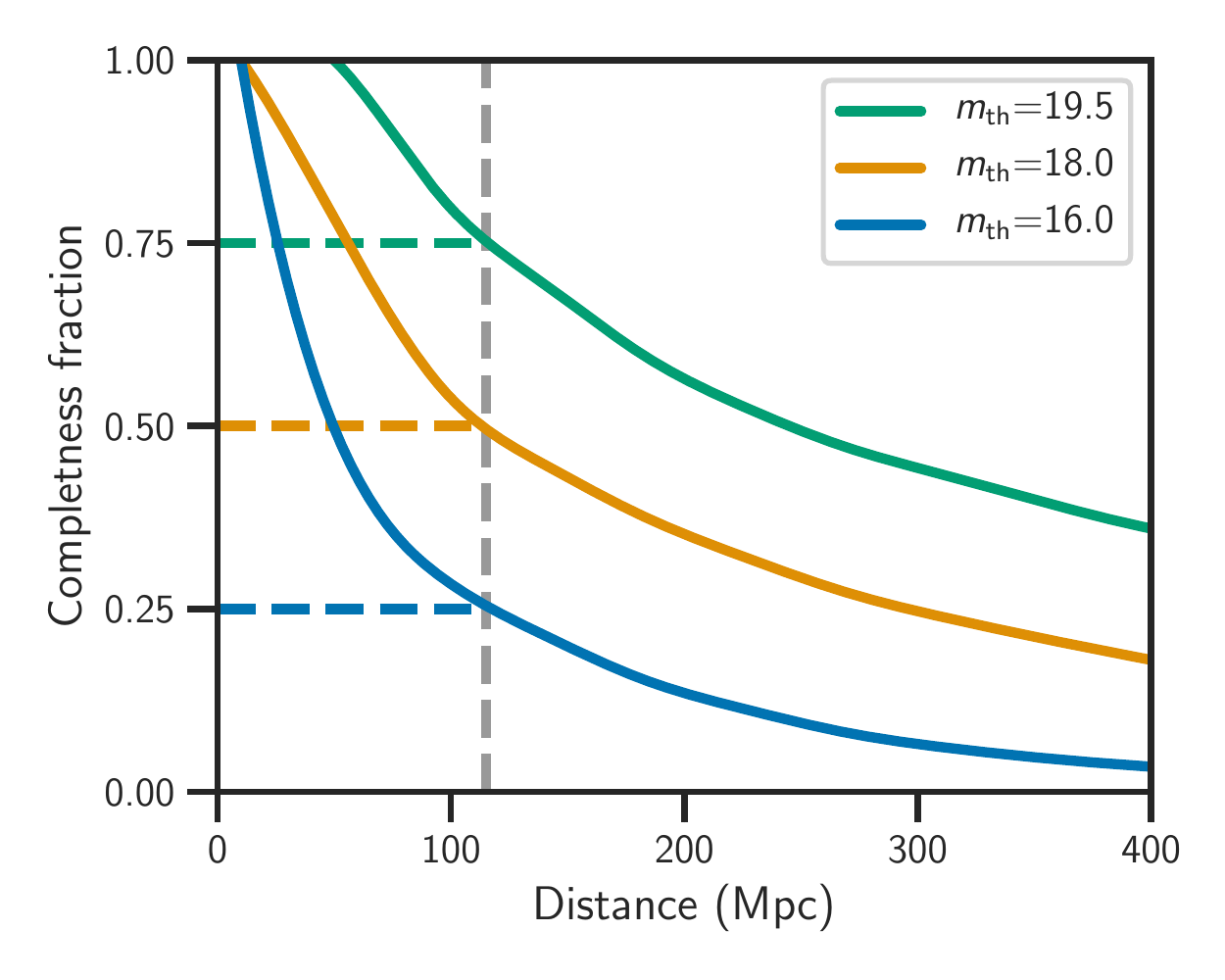}
\includegraphics[width=0.48\textwidth]{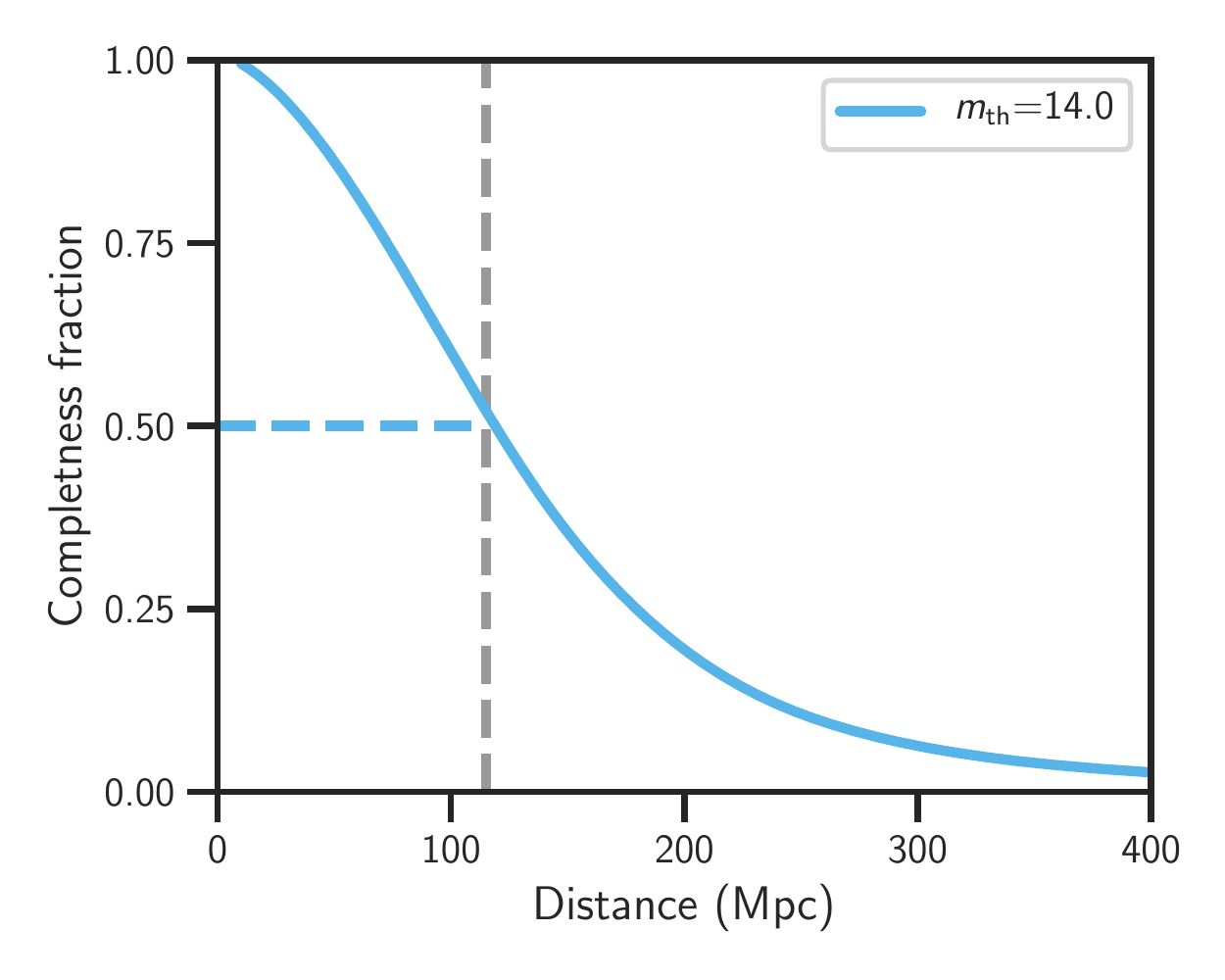}
\caption[Galaxy catalog completeness fractions for \ac{MDC}2 and 
\ac{MDC}3]{Galaxy catalog completeness fractions for \ac{MDC}2 and \ac{MDC}3. 
{\em Left panel:} Galaxy number completeness fraction defined in 
Eq.~\eqref{Eq:completeness} as a function of luminosity distance for the three 
\ac{MDC}2 sub-catalogs. The lines in green, orange and blue correspond to the 
catalogs with $m_\text{th}= 19.5$, $18$, and $16$ respectively; these correspond 
to completeness fractions of 75\%, 50\% and 25\% out to a fiducial reference 
distance of $115\,\Mpc$ (shown as a vertical grey line). {\em Right panel:} The 
galaxy luminosity completeness fraction defined in 
Eq.~\eqref{Eq:lumcompleteness} as a function of luminosity distance for the 
\ac{MDC}3 catalog, with $m_\text{th}=14$. At the reference distance of 115 Mpc 
(vertical grey line), this is corresponds to a completeness fraction of $\sim 
50\%$.}
\label{Fig:mdc2n3cum}
\end{figure*}

In this section we describe a series of mock data analyses (\acp{MDC}) that 
we use to test our implementation of the Bayesian formalism described in Section 
\ref{sec:methodology} and its ability to infer the posterior on \ac{H0} under 
different conditions. For each case, the \ac{MDC} consists of (i) simulated 
\ac{GW} data, and (ii) a corresponding mock galaxy catalog.
In all cases, we make several idealized assumptions regarding both the GW and galaxy data. On the GW side, the detection efficiency and the source population properties are assumed to be known exactly. On the galaxy side, the luminosity function and magnitude limit are also assumed to be known exactly in each case, so that the incompleteness correction can be calculated exactly. Further, we neglect the effects of large-scale structure and redshift uncertainties in the mock catalogs.

For each of the \acp{MDC} we use an identical set of simulated \ac{BNS} events from The First Two Years of Electromagnetic Follow-Up with Advanced LIGO and Virgo dataset \cite{2014ApJ...795..105S,Berry:2014jja}\footnote{The set of simulations in~\cite{Berry:2014jja} are more realistic with the same injections in (recolored) detector data as opposed to Gaussian noise used in~\cite{2014ApJ...795..105S}. Correspondingly, the detection criterion is in terms of a false alarm rate (FAR) rather than a threshold on the SNR. This is an important distinction, particularly affecting events marginally close to the detection threshold. We use the simplified set of simulations in~\cite{2014ApJ...795..105S} noting potential caveats.}.
The set of \ac{BNS} events comes from an end-to-end simulation of approximately 50,000 ``injected'' events in detector noise corresponding to a sensitivity similar to what was achieved during \ac{O2}. Only a subset (approximately 500 events) were ``detected'' by a network of two or three detectors with the \code{GstLAL} matched filter based detection pipeline \cite{gstlal:methods}. From the above detections, 249 events were randomly selected (in a way that no selection bias was introduced), and these events underwent full Bayesian parameter estimation using the \code{LALInference} software library \cite{LALInference} to obtain gravitational wave posterior samples and skymaps. Consistency with the First Two Years parameter estimation results in terms of sky localization areas and 3D volumes was demonstrated in~\cite{DelPozzo:2018dpu}. It is these 249 events of the First Two Years dataset and the associated \ac{GW} data which we use for our analysis.

The galaxy catalogs for each iteration of the \ac{MDC} described below are designed to test a new part of the \code{gwcosmo} methodology in a cumulative fashion, starting with \ac{GW} selection effects, adding in \ac{EM} selection effects, and finally testing the ability to utilize the information available in the observed brightness of host galaxies, by weighting the galaxies with a function of their intrinsic luminosities.

The starting point for the galaxy catalogs is to take all 50,000 injected events from the First Two Years dataset and simulate a mock universe, which contain a galaxy corresponding to each injected event's sky location and luminosity distance, where the latter is converted to a redshift using a fiducial ``simulated'' \ac{H0} value of $70$ \kmsMpc.  The First Two Years data was originally simulated in a universe where \ac{GW} events followed a $d_L^2$ distribution, and there was no distinction between the source frame and the (redshifted) detector frame masses. Though not ideal, this data reasonably mimics a low redshift universe ($z \ll 1$) in which the linear Hubble relation of Eq.~\eqref{Eq:Hubble} holds, and galaxies follow a $z^2$ distribution. We use the same linear relation for the generation of the \ac{MDC} universe ({\em i.e.} a set of simulated galaxy catalog parameters) for each of the \acp{MDC}. It should be emphasized that the Bayesian method for estimation of \ac{H0} outlined in Section~\ref{sec:methodology} above is general, and can be extended to realistic scenarios with a non-linear cosmology with $\{ \Omega_\text{m}, \Omega_\text{k}, \Omega_\Lambda\}$ held fixed. So, in particular, the method is applicable for events which are detected at higher distances, where the low redshift approximation breaks down. The restriction to a linear cosmology in this paper comes only due to the use of the \ac{MDC} dataset. We would like to note that by using a linear cosmology, we are not testing possible effects introduced by the presence of other cosmological parameters. 
The analysis at large redshifts may, for example, be sensitive to the values (or the assumed prior ranges) of the parameters like $\Omega_\text{m}$ and $\Omega_\Lambda$.

The first four columns of Table \ref{tab:results} summarize the characteristics of each of the galaxy catalogs created and how they correspond to each \ac{MDC}. We give a brief description for each of the cases below.

\subsection{\ac{MDC}0: Known Associated Host Galaxies}
\ac{MDC}0 is the simplest version of the \acp{MDC}, in which we identify with certainty the host galaxy for each \ac{GW} event, and is equivalent to the direct counterpart case. As the galaxies are generated with no redshift uncertainties or peculiar velocities, the results will be (very) optimistic. This \ac{MDC} provides the ``best possible'' constraint on \ac{H0} using the 249 events, which then allows for comparison with the other \acp{MDC}.

\subsection{\ac{MDC}1: Complete Galaxy Catalog}
The \ac{MDC}1 universe consists of the full set of 50,000 galaxies out to $z\approx0.1$ ($d_L\approx428\,\Mpc$) in the original First Two Years dataset. This gives a galaxy number density of $\sim$ 1 per $7000\,\Mpc^3$, which is $\sim35$ times sparse compared to the actual density of galaxies in the local universe~\cite{GLADE}. Additional galaxies are generated beyond the edge of the dataset universe, uniformly across the sky and uniformly in comoving volume, thereby extending the universe out to a radius of 2000 Mpc ($z=0.467$ for $H_0=70$ \kmsMpc). This means that, even allowing \ac{H0} to be as large as 200 \kmsMpc, the edge of the \ac{MDC} universe is more than twice the highest redshift associated with the farthest detection (which is at $\sim 270$ Mpc)\footnote{For \ac{MDC}1 and for all subsequent \acp{MDC}, it has been tested that the artificial ``edge of the universe'' has no bearing on the results.}. Each of the 249 detected \ac{BNS} have a unique associated host galaxy contained within the \ac{MDC}1 catalog. This catalog is thus \emph{complete} in the sense that it contains every galaxy in the simulated universe. We refer to the \ac{MDC} universe as \ac{MDC}1 throughout the rest of the paper, and similarly for the subsequent \acp{MDC}.

\ac{MDC}1 is designed to test our treatment of \ac{GW} selection effects, by ensuring that given a set of sources and access to a complete catalog, our methodology and analysis produces a result consistent with the simulated value of \ac{H0}.

\subsection{\ac{MDC}2: Incomplete Galaxy Catalog \label{sec:MDC2}}

\ac{MDC}2 is designed to test our treatment of \ac{EM} selection effects, by applying an apparent magnitude threshold to the \ac{MDC} universe, such that a certain fraction of the host galaxies is not contained in it. This is a necessary consideration, given that we are in the regime where \ac{GW} signals are being detected beyond the distance to which the current galaxy catalogs can be considered to be complete. This has been true for \acp{BBH} detections since \ac{O1}, and is true of \acp{BNS} as well in \ac{O3}.

In order to create the catalog for \ac{MDC}2, we start with the initial \ac{MDC}1 universe and assign luminosities to each of the galaxies within it. We assume that the luminosity distribution of the galaxy catalog is known to the observer throughout and
follows a Schechter function of the form~\cite{Schechter:1976}
\begin{equation}\label{Eq:Schechter}
\phi(L) \, dL = n^* \left(\frac{L}{L^*}\right)^\alpha \ e^{-L/L^*} \ \frac{dL}{L^*} \,,
\end{equation}
where $L$ denotes a given galaxy luminosity and $\phi(L)\,dL$ is the number of 
galaxies within the luminosity interval $[L, L + dL]$. The characteristic galaxy 
luminosity is given by $L^* = 1.2 \times 10^{10} \, h^{-2} \, L_\odot$ with 
solar luminosity $L_\odot = 3.828\times 10^{26}\,\mathrm{W}$, and $h\equiv 
H_0/(100\kmsMpc)$\footnote{We note that the parameter $L^*$ of the Schechter 
luminosity function itself depends on $H_0$, which we allow to vary and hence 
take into account within our formalism.},
$\alpha = -1.07$ characterizes the exponential drop off of the luminosity function,
and $n^*$ denotes the number density of objects in the \ac{MDC} universe (in practice, this only acts as a normalization constant). The integral of the Schechter function diverges at $L \rightarrow 0$, requiring a lower luminosity cutoff for the dimmest galaxies in the universe which we set to $L_{\text{lower}} = 0.001L^*$. This choice is arbitrary for our purpose here, but small enough to include almost all objects classified as galaxies in real catalogs like GLADE~\cite{GLADE}.

These luminosities are then converted to apparent magnitudes using $m\equiv25-2.5\log_{10}(L/L^*)+5\log_{10}(d_L/\Mpc)$, and an apparent magnitude threshold $m_\text{th}$ is applied as a crude characterization of the selection function of an optical telescope observing only objects with $m<m_\text{th}$. \ac{MDC}2 is broken into three sub-\acp{MDC}, in order to test our ability to handle different levels of galaxy catalog completeness dictated by different telescope sensitivity thresholds. In each case, the catalog completeness is defined as the ratio of the number of galaxies inside the catalog relative to the number of galaxies inside the \ac{MDC} universe, out to a reference fiducial distance $d_L$,
\begin{equation}\label{Eq:completeness}
f_\text{completeness}(d_L) = \dfrac{\sum_j^\text{MDA2} \Theta(d_L - d_{L_j})}{\sum_k^\text{MDA1} \Theta(d_L - d_{L_k})}\,,
\end{equation}
where the numerator is a sum over the  galaxies contained within the \ac{MDC}2 catalog out to some reference distance $d_L$, and the denominator is a sum over the galaxies in the \ac{MDC}1 catalog.

Apparent magnitude thresholds of $m_\text{th}=19.5$, $18$, and $16$ are chosen for the three sub-\acp{MDC}, which correspond to cumulative number completeness fractions of $~75\%$, $~50\%$ and $~25\%$ respectively, evaluated at a distance of $d_L = 115\, \Mpc$, chosen such that given the luminosity distance distribution of detected \acp{BNS}, the completeness fraction for the sub-\ac{MDC} to this distance is roughly indicative of the percentage of host galaxies which remain inside the galaxy catalog. The left panel of Fig.~\ref{Fig:mdc2n3cum} shows how the completeness of each of the \ac{MDC}2 catalogs drop off as a function of distance.

\subsection{\ac{MDC}3: Luminosity Weighting}\label{ss:MDC3}

\ac{MDC}3 is designed to test the effect of weighting the likelihood of any galaxy being host to a \ac{GW} event as a function of their luminosity. It is probable that the more luminous galaxies are also more likely hosts for compact binary mergers; the luminosity in blue (B-band) is indicative of a galaxy's star formation rate, for example, while the luminosity in high infrared (K-band) is a tracer of the stellar mass~\cite{Singer:2016,Leibler:2010,Fong:2013}. The bulk of the host probability is expected to be contained within a smaller number of brighter galaxies, effectively reducing the number of galaxies which need to be considered. Additional information from luminosity is thus expected to improve the constraint on \ac{H0} by narrowing its posterior probability density.

For \ac{MDC}3, the probability of a galaxy hosting a \ac{GW} event is
chosen to be proportional to the galaxy's luminosity.  Because the
\ac{GW} events for these \acp{MDC} were generated in advance, and we are
retroactively simulating the universe in which they exist, generating the
\ac{MDC}3 universe required some care: luminosities have to be assigned to the
host galaxies \emph{and} the non-host galaxies in such a way that our choice of
simulated luminosity weighting is correctly represented within the galaxy
catalog.

As with \ac{MDC}2, the luminosity distribution of the galaxies in the universe is assumed to follow the Schechter luminosity function as in Eq.~\eqref{Eq:Schechter} (referred to from now on as $p(L)$).  However, the joint probability of a single galaxy having luminosity $L$ and hosting a \ac{GW} event (which emits a signal, $s$) is $p(L,s) \propto L\,p(L)\,,$
where we assume that the probability of a galaxy of luminosity $L$ hosting a source is proportional to the luminosity itself
. All host galaxies thus have luminosities sampled from $L\,p(L)$. In this context, we must consider all galaxies which hosted \ac{GW} events, not just those from which a signal was detected.
With this in mind, the overall luminosity distribution has the following form:
\begin{equation}
p(L) = \beta \dfrac{L}{\langle L \rangle}p(L) + (1-\beta)\,x(L)
\end{equation}
where $\beta$ is the fraction of host galaxies to total galaxies for the observed time period ($1 \geq \beta \geq 0$), $L/ \langle L \rangle$ is the normalized luminosity, and $x(L)$ is the unknown luminosity distribution of galaxies which did not host \ac{GW} events, which we can sample for a given value of $\beta$. 

Rearranging to obtain the only unknown, $x(L)$, gives
\begin{equation}
x(L) = \dfrac{p(L)}{1-\beta} \Bigg[1 - \beta\dfrac{L}{\langle L \rangle}\Bigg]\,,
\end{equation}
and from this we see there is an additional constraint on $\beta$, because the term inside the brackets must be $>0$.  The maximum value that $\beta$ can take is given by $\beta_{\text{max}} = \langle L \rangle/L_{\text{max}}$, where $L_{\text{max}}$ is the maximum luminosity from the Schechter function, and $\langle L \rangle$ is the mean.  From the Schechter function parameters detailed in section \ref{sec:MDC2}, $\beta_{\text{max}} \approx 0.015$.

The original First Two Years data was generated by simulating $\sim 50,000$ \ac{BNS} events, of which $\sim 500$ were detected, of which 249 randomly selected detections underwent parameter estimation. The number of ``hosting'' and ``non-hosting'' galaxies have to be rescaled to represent this. Thus half of the original galaxies were denoted as hosts (including those associated with the 249 detected \ac{GW} events). However, in order to satisfy the requirements for $\beta$, a greater density of non-hosting galaxies had to be added to the universe before luminosities could be assigned. Thus for \ac{MDC}3, the density of galaxies is increased by a factor of 100, with the acknowledgement that this would lead to a broadening of the final posterior.  
\ac{MDC}3 has a galaxy density of $\sim1$ galaxy per $70\,\Mpc^3$, which is about $3$ times denser than the actual density of galaxies in the local universe~\cite{GLADE}.
This also means that \ac{MDC}3 is not directly comparable with the previous \ac{MDC} versions, save \ac{MDC}0.  The galaxies which are hosts are assigned luminosities from $Lp(L)$, and non-hosts from $x(L)$ above.

In order to include \ac{EM} selection effects, an apparent magnitude cut $m_\text{th}$ of 14 is applied, such that the completeness of the galaxy catalog is $\sim 50\%$ out to the same fiducial distance of $115\, \Mpc$ as in \ac{MDC}2. In this case, completeness is however defined in terms of the fractional luminosity contained in the catalog, rather than in terms of numbers of objects:
\begin{equation}\label{Eq:lumcompleteness}
f_\text{completeness}(d_L) = \dfrac{\sum_j^\text{MDA3} L_j \Theta(d_L-d_{L_j})}{\sum_k^\text{complete} L_k \Theta(d_L-d_{L_k})}\,,
\end{equation}
where the numerator is summed over the galaxies inside the \ac{MDC}3 apparent magnitude-limited catalog, and the denominator is summed over the galaxies in the whole \ac{MDC}3 universe. This is shown in the right panel of Fig.~\ref{Fig:mdc2n3cum}. As the host galaxies are luminosity weighted, the cumulative luminosity completeness is representative of the percentage of \ac{BNS} event hosts inside the catalog.

\section{Results\label{sec:results}}
{\renewcommand{\arraystretch}{1.5}
\begin{table*}[ht]
\centering
\begin{ruledtabular}
\begin{tabular}{ccccccc}
\ac{MDC} & Host galaxy preference & Completeness\footnote{The completeness is 
calculated as a number completeness using Eq.~\eqref{Eq:completeness} for 
\acp{MDC} 1 and 2, and as a luminosity completeness using 
Eq.~\eqref{Eq:lumcompleteness} for \ac{MDC} 3, out to a fiducial distance of 115 
Mpc, such that it is indicative of the fraction of host galaxies which are 
inside the galaxy catalog in both cases.} & $m_{\text{th}}$ & Analysis 
assumption & $H_0$ (\kmsMpc) & Fractional uncertainty \\ \hline \hline
0 & Known host & - & - & direct counterpart & \Hubblemdccounter & 
\Hubblemdccountererror \% \\ \hline
1 & equal weights & 100\% & - & unweighted catalog & \Hubblemdc & 
\Hubblemdcerror \% \\ \hline
2a & equal weights & 75\% & 19.5 & unweighted catalog & \Hubblemdcc & 
\Hubblemdccerror \% \\ \hline
2b & equal weights & 50\% & 18 & unweighted catalog & \Hubblemdcb & 
\Hubblemdcberror \% \\ \hline
2c & equal weights & 25\% & 16 & unweighted catalog & \Hubblemdca & 
\Hubblemdcaerror \% \\ \hline \hline
3a & luminosity weighted & 50\% & 14 & weighted catalog & \Hubblemdcweights & 
\Hubblemdcweightserror \% \\ \hline
3b & luminosity weighted & 50\% & 14 & unweighted catalog & \Hubblemdcnoweights 
& \Hubblemdcnoweightserror \% \\ 
\end{tabular}
\end{ruledtabular}
\caption{\label{tab:results} A summary of the main results. We quote the peak value and the 68.3\% highest density error region for the posterior probability on $H_0$ for each of the \acp{MDC} combining all 249 events. The fractional uncertainty is defined as the half-width of the 68.3\% highest density probability interval divided by the simulated value of $H_0=70$ \kmsMpc.}
\end{table*}

In this section we summarize the results for the mock data analyses described 
in Section \ref{sec:mdc}. We show the combined posteriors on \ac{H0} for each 
\ac{MDC}, discuss the convergence to the simulated value of \ac{H0} = $70$ 
\kmsMpc and calculate the precision of the combined measurement under 
each set of conditions. In Table \ref{tab:results} we list the measured values 
of the Hubble constant for the combined 249 event posterior (maximum a-posteriori and 68.3\% highest density posterior intervals) all computed with a 
uniform prior on $H_0$ in the range of \priorrange, as well as the corresponding 
fractional uncertainties for each of the \acp{MDC}.

\subsection{\ac{MDC}0: Known Associated Host Galaxies}

\begin{figure*}
\includegraphics[width=\linewidth]{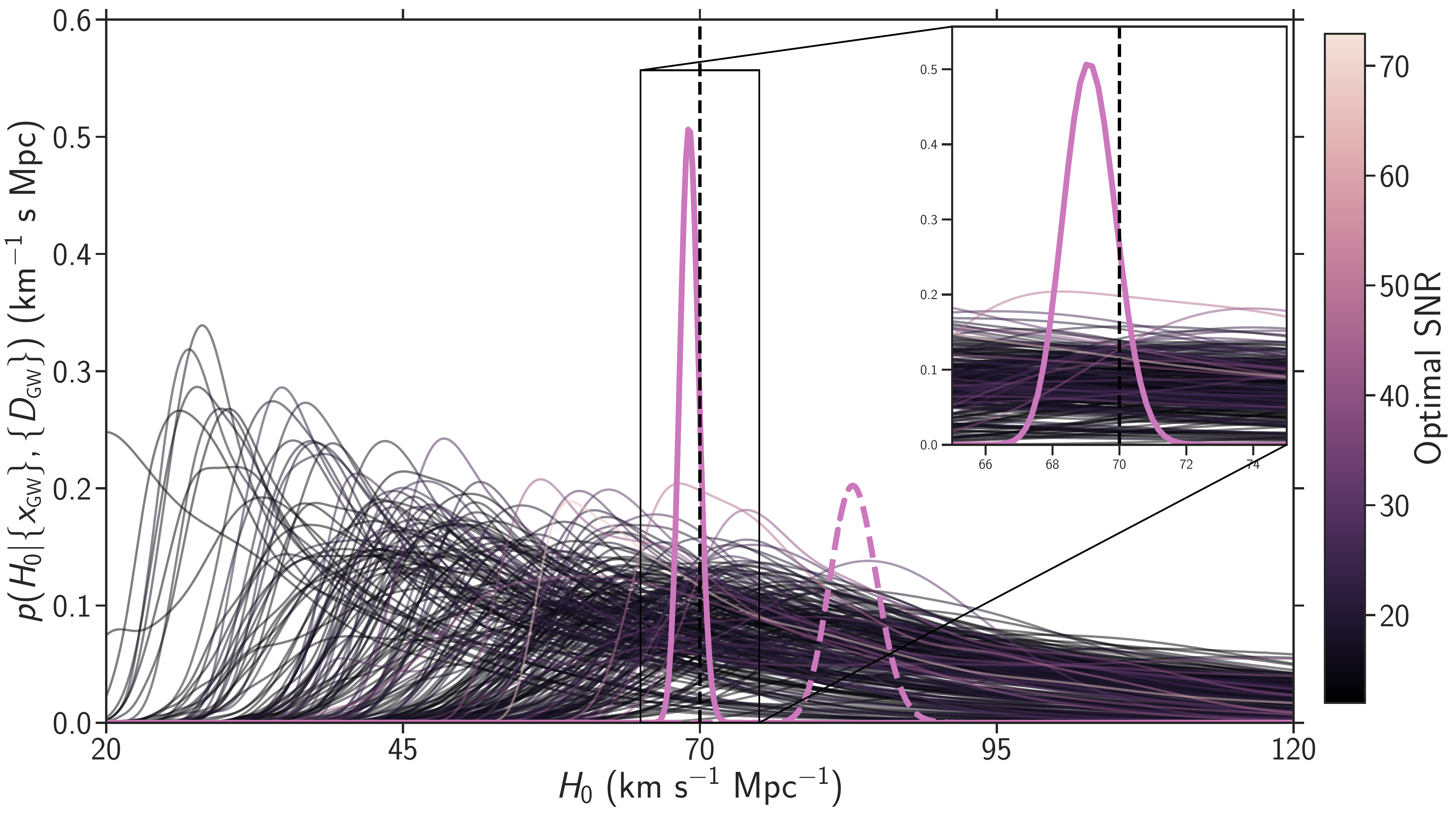}
\caption{Individual and combined results for \ac{MDC}0 (known host galaxy or direct counterpart case). The solid thick purple line shows the combined posterior probability density on $H_0$, while the dashed line shows the combined posterior when \ac{GW} selection effects are neglected. Individual likelihoods (normalized and then scaled by an arbitrary value), for each of the 249 events, are shown as thin lines with shades corresponding to their optimal SNR. The simulated value of $H_0$ is shown as a vertical dashed line.}
\label{fig:mdc0all}
\end{figure*}

We first consider the simple case where we identify the true host galaxy for 
every event and determine the resulting 249-event combined $H_0$ posterior.
Fig.~\ref{fig:mdc0all} presents the results of this analysis.
The likelihoods for each individual \ac{GW} event are shown
(normalized relative to each other but scaled with respect to the combined posterior
for clarity) shaded by the 
event's optimal SNR in the detector network, as defined in \cite{Cutler_1994}.
In this case, each likelihood is informative, having a clearly-defined peak 
corresponding to finding the likely values of $H_0$ for the known galaxy redshift.
Each curve traces the information in the corresponding $d_L$ distribution,
which is usually unimodal, but in some cases 
may have two or more peaks~\cite{2014ApJ...795..105S,Berry:2014jja}. We see that the peaks of 
the individual likelihoods do not necessarily correspond to the true value 
$H_0=70\,\kmsMpc$, but there is always support for it, leading to the combined posterior, which
is overlaid in thick purple. This gives us a statistical estimate for the 
maximum a-posteriori value and $68.3\%$ maximum-density credible interval for 
$H_0$ as $\Hubblemdccounter \, \kmsMpc$.
The final result combining all the 249 events have converged
to a relatively symmetric ``Gaussian'' distribution~\cite{Walker_1969}.

The result of \ac{MDC}0 provides us with the best possible \ac{H0} estimate 
given the set of \ac{GW} detections, since this case corresponds to perfect 
knowledge of the host galaxies. This gives us a benchmark against 
which other versions of the \ac{MDC} can be compared.
Since this is a best-case scenario, we have the least statistical uncertainty
in the final result, making any systematic bias more apparent than for the subsequent
\acp{MDC}. 
For the combined result with 249 events, the simulated value is contained within the 
support of the posterior distribution of $H_0$.

\ac{MDC}0 demonstrates the importance of correctly accounting for
\ac{GW} selection effects. We are biased towards detecting sources which are
nearer-by, and which are optimally orientated (closer to face-on).  If an
analysis is performed without taking into consideration the denominator
$p(D_\text{GW}|H_0)$ of Eq.~\eqref{Eq.xD}, which corrects for this, the
posterior density on $H_0$ converges to a value different from its simulated
value of $70\,\kmsMpc$.  This can be seen in Fig. \ref{fig:mdc0all}, where the
dashed purple line shows the \ac{MDC}0 combined posterior for all 249 events,
neglecting \ac{GW} selection effects entirely. We leave a detailed
exploration of what level of accuracy in the \ac{GW} selection function is
required in order to move beyond 249 \ac{BNS}-with-counterpart events, and
simply note that in this case, it is sufficient enough that any biases which
could affect the next stages of the \ac{MDC} do not arise from the \ac{GW}
selection effects.

\subsection{\ac{MDC}1: Complete Galaxy Catalog}

\begin{figure}
\includegraphics[width=\linewidth]{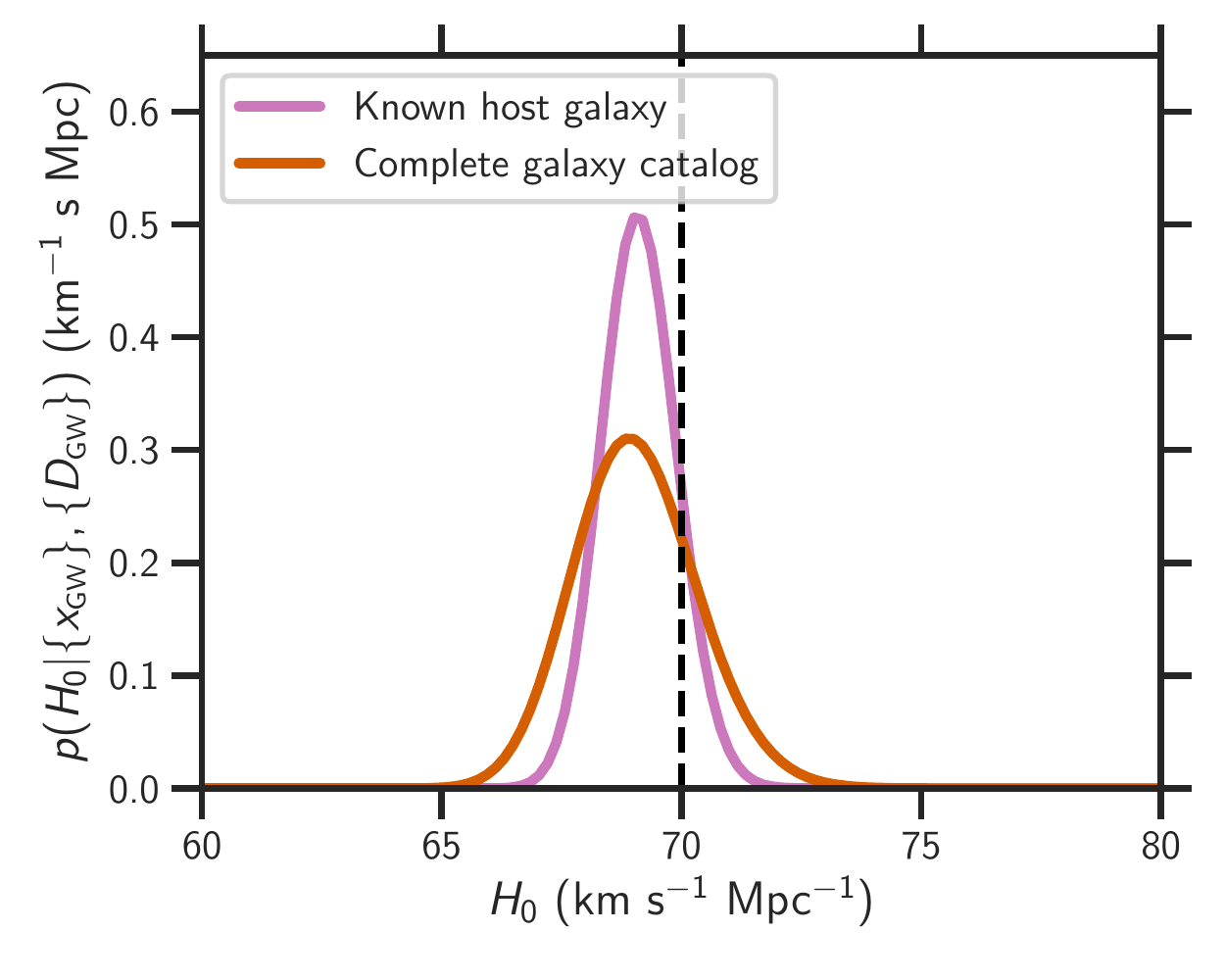}
\caption{Comparison of the galaxy catalog method with the known host galaxy case. Joint posterior probability density on $H_0$ using all 249 events for \ac{MDC}0 (known host galaxy) and \ac{MDC}1 (complete galaxy catalog) are shown respectively in purple and red. For this set of simulations, uncertainty with the galaxy catalog is only about $1.63$ times larger than with known host galaxies.}
\label{fig:mdc01}
\end{figure}

\begin{figure*}
\includegraphics[width=\linewidth]{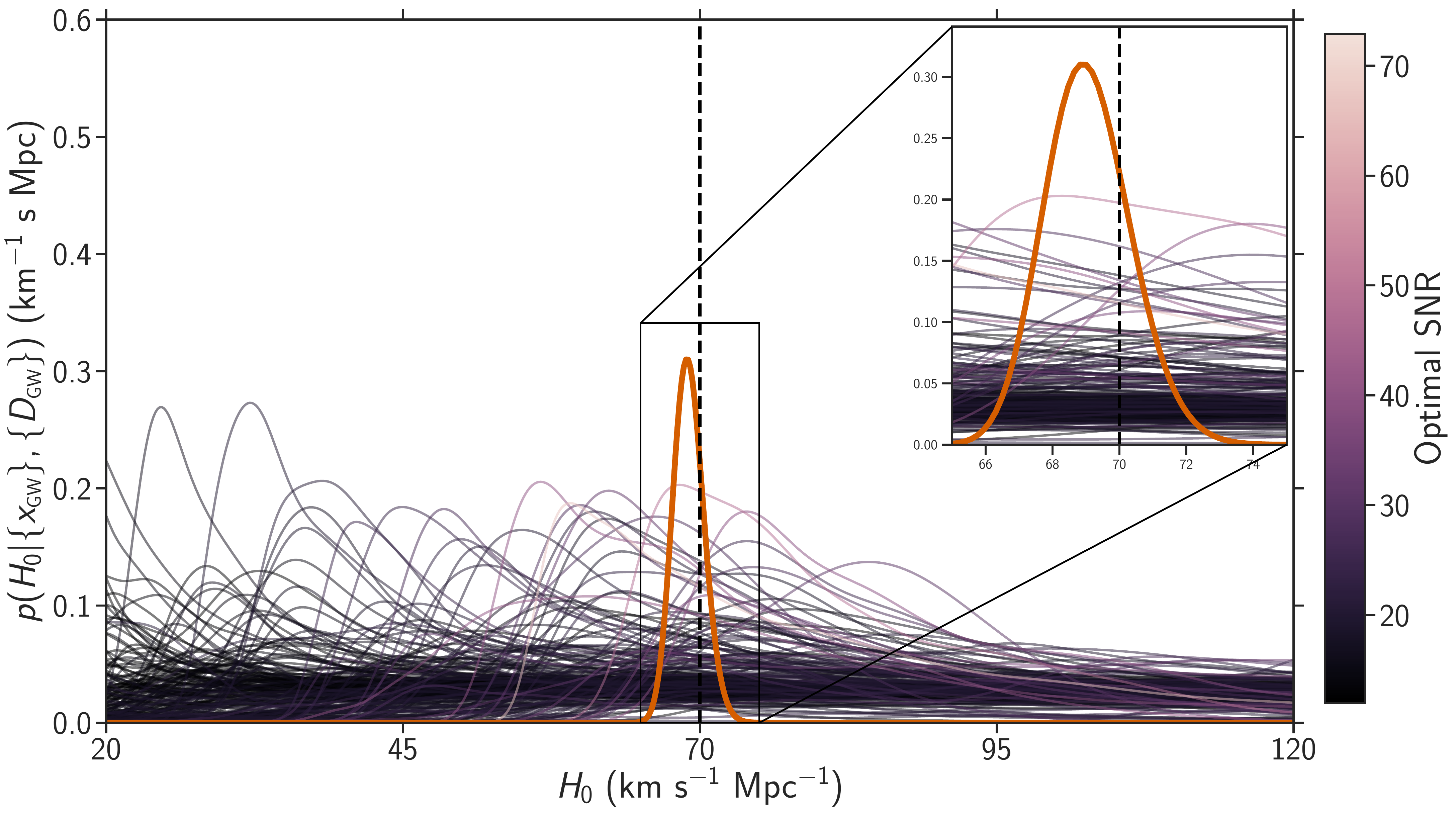}
\caption{Individual and combined results for \ac{MDC}1 (complete galaxy catalog). The thick red line shows the combined posterior probability density on $H_0$. Individual likelihoods (normalized and then scaled by an arbitrary value), for each of the 249 events, are shown as thin lines with shades corresponding to their optimal SNR. The simulated value of $H_0$ is shown as a vertical dashed line. Many of the individual likelihoods do not have sharp features, however the final result  converges to the simulated value with redshift information present in the galaxy catalogs. This demonstrates the applicability of our methodology.}
\label{fig:mdc1all}
\end{figure*}

The next more complex case is \ac{MDC}1, where we assume no counterpart was 
observed, and resort to using a galaxy catalog. \ac{MDC}1 uses
a \emph{complete} galaxy catalog containing all potential hosts -- an optimistic scenario, in which EM selection effects do not need to be considered.
The results with
\ac{MDC}1 already show a wider posterior distribution on \ac{H0} ($\Hubblemdc\,\kmsMpc$)
because of lack of certainty of the host galaxy (Fig.~\ref{fig:mdc01}).
The introduction of this uncertainty means that the contributions from each 
event will be smoothed out, depending on the size of the event's sky 
localization and the number of galaxies within it. As can be seen in
Fig.~\ref{fig:mdc1all}, there is a far higher proportion of events for which the 
likelihood is relatively broad and less informative, in comparison to
Fig.~\ref{fig:mdc0all}.  However, many events clearly have a small number of 
galaxies in their sky-area, and hence still show clear peaks.

\subsection{\ac{MDC}2: Incomplete Galaxy Catalog}\label{ss:MDC2}
\begin{figure}
\includegraphics[width=\linewidth]{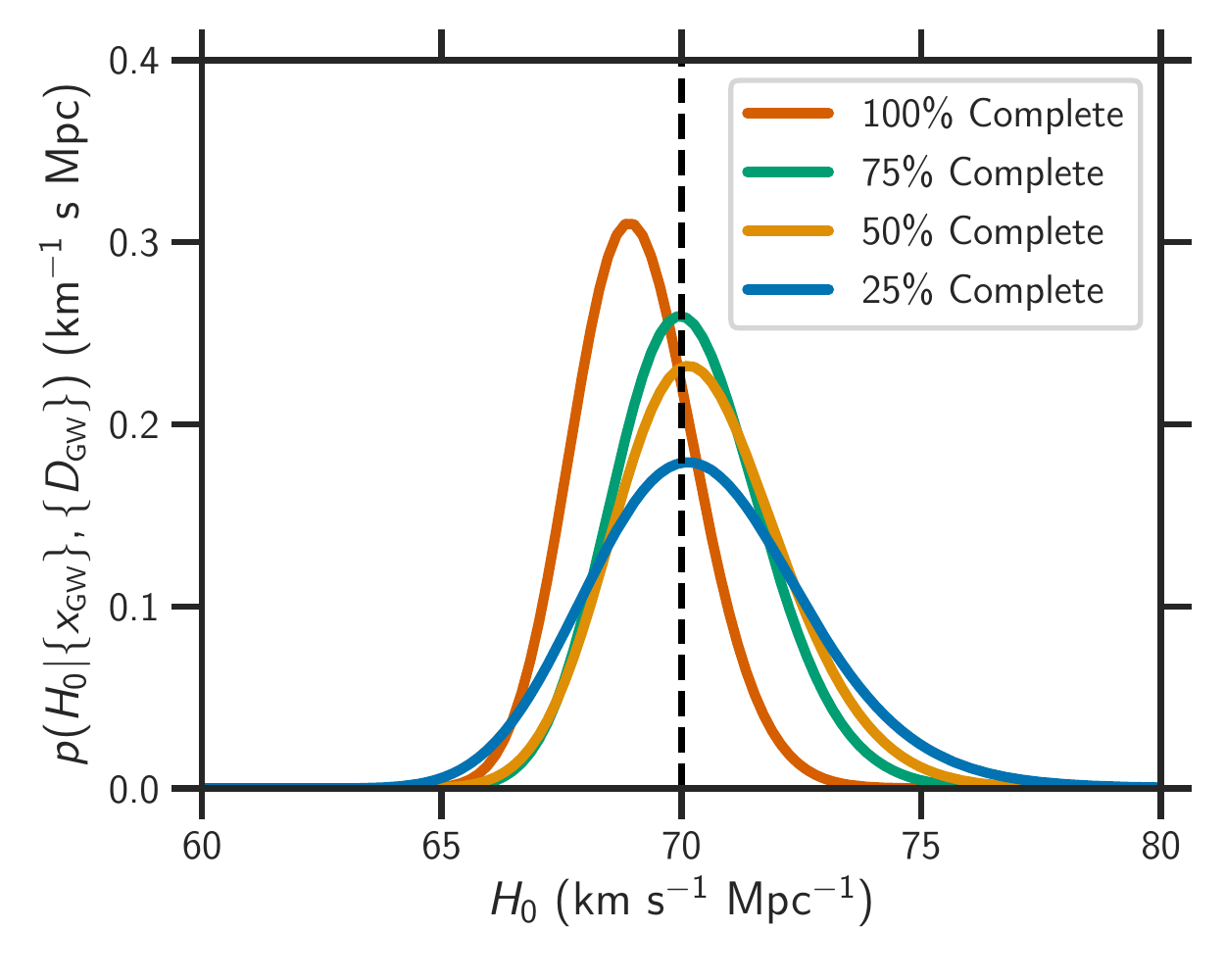}
\caption{\label{fig:mdc2result} Comparison of results with varying galaxy catalog completeness. In \ac{MDC}2, the simulated apparent magnitude threshold is varied to obtain galaxy catalogs of $100\%$, $75\%$, $50\%$, and $25\%$ completeness. The corresponding posterior probability densities on $H_0$ using all 249 events are shown in red, green, yellow, and blue respectively.}
\end{figure}

The next most complex scenario is the case where we have incomplete galaxy 
catalogs, limited by an apparent magnitude threshold.
This gives us the first case where accounting for EM selection effects is
important. To investigate this, we consider three galaxy catalogs, with
apparent magnitude thresholds of $m_\text{th}=19.5$, $18$ and $16$, with respective completeness 
fractions of $75\%$, $50\%$ and $25\%$ in addition to the complete catalog
for \ac{MDC}1 (see \ref{sec:MDC2} for details).
The combined 249-event posterior distributions on \ac{H0} are shown in 
Fig.~\ref{fig:mdc2result}.

\begin{figure*}
\includegraphics[width=\linewidth]{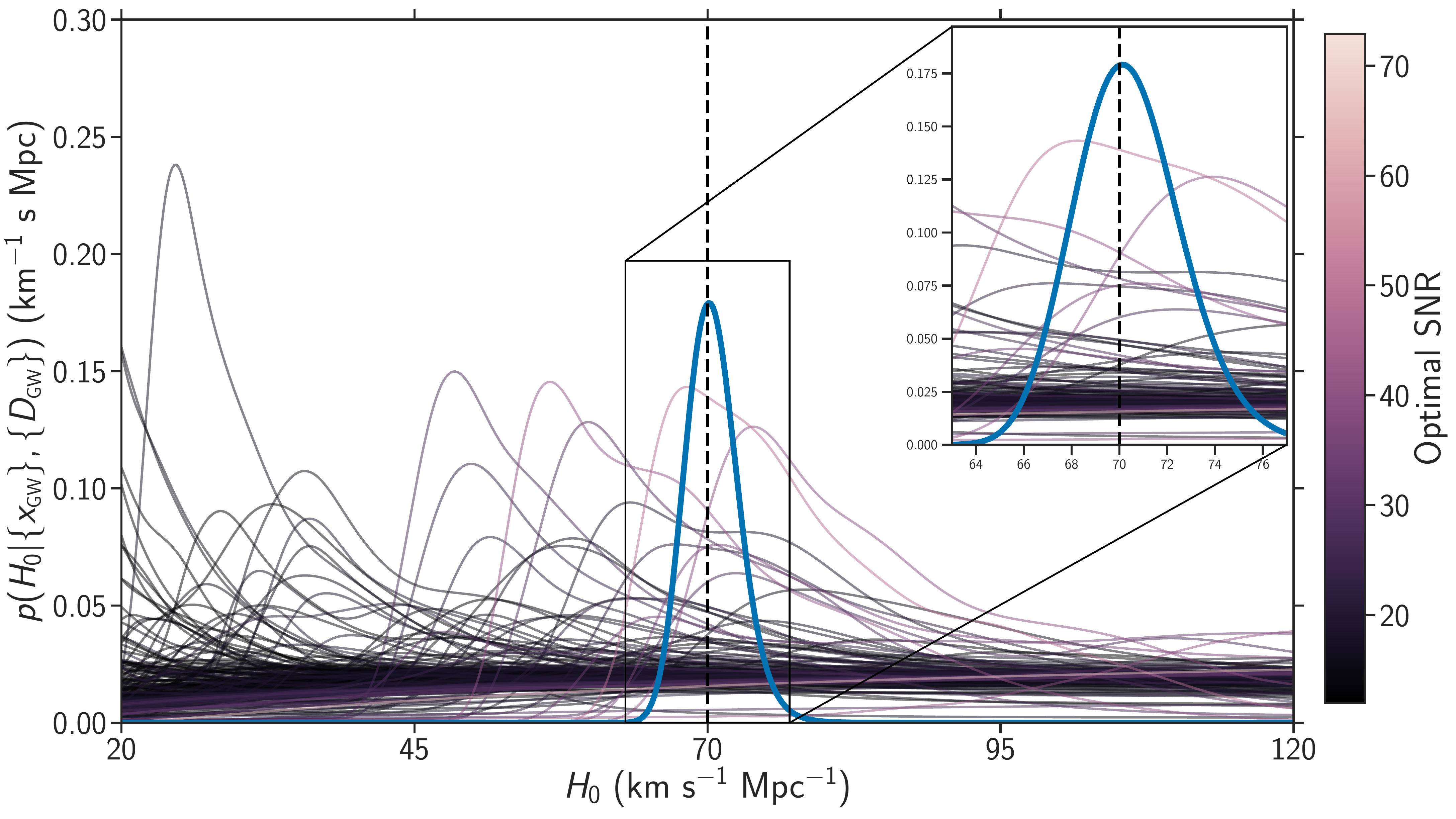}
\caption{\label{fig:mdc_25_all}Individual and combined results for \ac{MDC}2 with a $25\%$ complete galaxy catalog. The thick blue line shows the combined posterior probability density on $H_0$. Individual likelihoods (normalized and then scaled by an arbitrary value), for each of the 249 events, are shown as thin lines with shades corresponding to their optimal SNR. The simulated value of $H_0$ is shown as a vertical dashed line. Compared to \ac{MDC}0 (Fig.~\ref{fig:mdc0all}) and \ac{MDC}1 (Fig.~\ref{fig:mdc1all}), fewer individual likelihoods are peaked here. Although the final $H_0$ estimate is less precise, the results converge to the simulated value, demonstrating the applicability of our methodology to threshold-limited galaxy catalogs of about $25\%$ completeness.
}
\end{figure*}

As the catalogs become less complete, the 
combined \ac{H0} posterior becomes wider. This is because the probability that 
the host galaxy is inside the catalog decreases. The 
contribution from the galaxies within the catalog is reduced, and the 
uninformative contribution from the out-of-catalog term in Eq.~\eqref{Eq:sum G} 
increases. This is visible in the individual likelihoods shown in 
Fig.~\ref{fig:mdc_25_all}, where instead of decreasing toward zero at high
values of $H_0$, many of the individual likelihoods tend toward a constant.
This is because, in the absence of EM data, and with the linear Hubble
relation assumed in this work, the number of unobserved galaxies increases
without limit as $d_L^2$.
This is seen mostly for events at high distances (where the host has a lower probability of being in the catalog), or for well-localized events where there is no catalog support at the relevant redshifts within the event's sky area.  
However, enough events are detected at low distances, where the catalogs are more complete and so provide informative redshift information, to produce an upper constraint on $H_0$.

We estimate $H_0=\Hubblemdcc$, $\Hubblemdcb$, 
and $\Hubblemdca$ $\kmsMpc$ 
respectively for galaxy catalogs
of 75\%, 50\%, and 25\% completeness. See 
section \ref{sec:mdc comparison} for a 
more in depth comparison of how galaxy catalog completeness affects posterior 
width.

Our exercise demonstrates that we need to {know} (or assess) the completeness of 
galaxy catalogs, and put in an appropriate out-of-catalog term in the analysis.
For any of the \ac{MDC}2 catalogs, if we assume that the galaxy catalog is 
complete, when in reality it is not, we get a posterior distribution on \ac{H0} which is 
inconsistent with a value of 70 \kmsMpc. 
This is because the well-localized events for 
which the host is not inside the catalog
do not have support for the correct value of $H_0$.
In real catalogs, galaxy clustering might ensure that there are nearby bright galaxies
in the catalog, partially mitigating this bias.

\subsection{\ac{MDC}3: Luminosity Weighting}
\begin{figure}
\includegraphics[width=\linewidth]{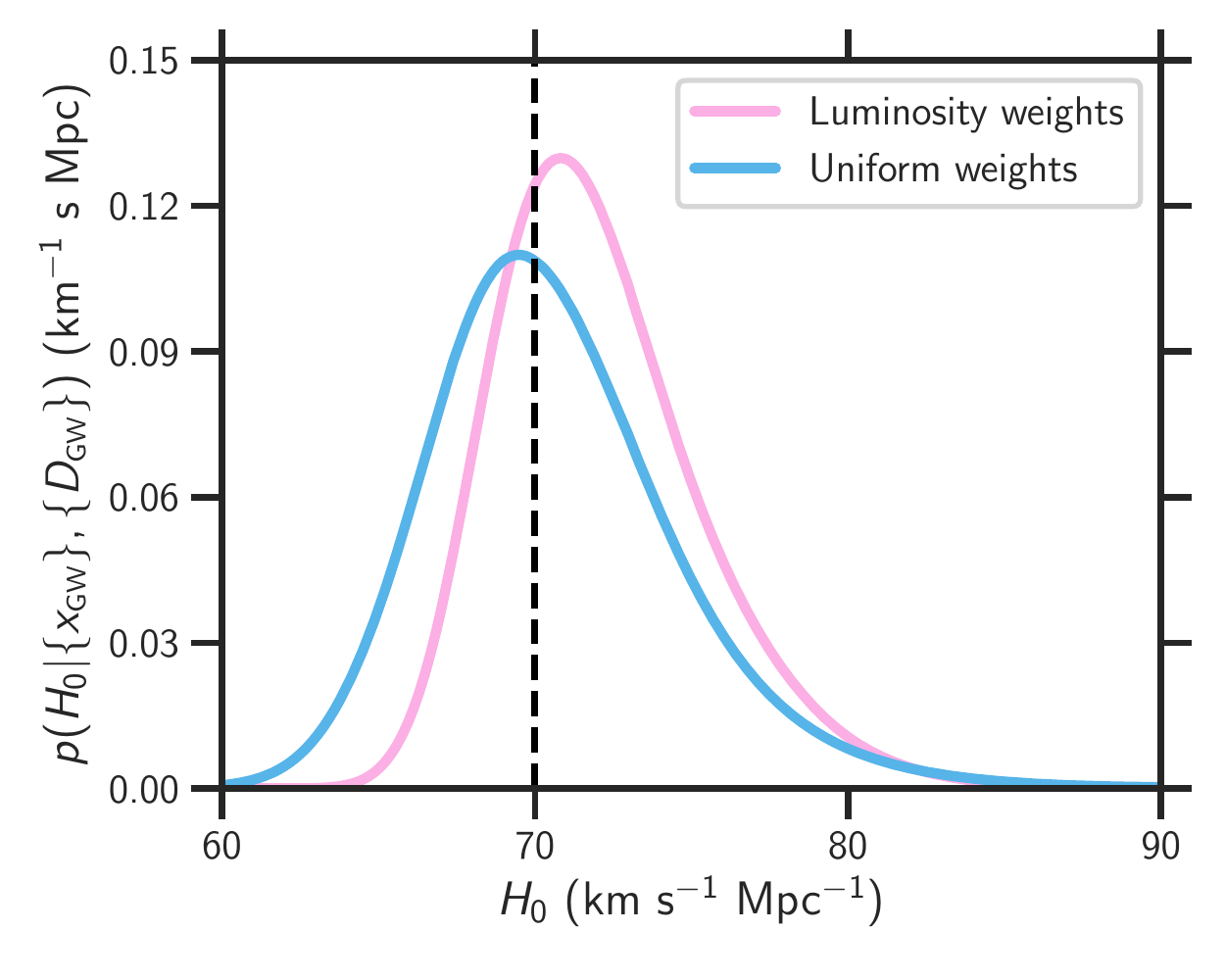}
\caption{\label{fig:mdc3result} Comparison of results with and without luminosity weighting. In \ac{MDC}3, by construction, the probability of any galaxy hosting a \ac{GW} event is proportional to its luminosity. The pink curve shows the posterior probability density on $H_0$ for the case where we take this into account in our analysis as a weighting by the galaxy's luminosity. The blue curve shows the posterior density for the case where we ignore this extra information, and treat every galaxy as equally likely to be hosts. Luminosity weighting improves the precision in the results by a factor of $1.2$ for this set of simulations.}
\end{figure}

Until now we have considered all galaxies in our catalog to be equally likely
to host a gravitational-wave source. In \ac{MDC}3 we analyze the 
case described in Sec.~\ref{ss:MDC3} where this is no longer true by constructing a galaxy catalog such that 
the probability of any single galaxy hosting a \ac{GW} source is directly 
proportional to its luminosity. \ac{MDC}3 includes the same \ac{EM} 
selection effects as \ac{MDC}2, in the sense that the catalog is magnitude limited.  The 
completeness of this catalog, defined in terms of luminosity rather than 
numbers of galaxies, as defined in Eq.~\eqref{Eq:lumcompleteness}, is 50\% out to 
115 Mpc.  This is indicative that approximately 50\% of the detected \ac{GW} 
events have host galaxies inside the catalog.

To investigate the importance of luminosity weighting, \ac{MDC}3 was analyzed twice 
under different assumptions, given in Eq.~\eqref{Eq:luminosityweighting}.  In 
the first, the analysis was matched to the known properties of the galaxy 
catalog, such that the probability of any galaxy hosting a \ac{GW} event was 
proportional to its luminosity. In the second, we feigned ignorance and ran the 
analysis with the assumption that each galaxy was equally likely to be host to a 
\ac{GW} event (as was true in \acp{MDC} 1 and 2). This allows us to determine 
the effect of ignoring galaxy weighting with this dataset.
The combined \ac{H0} posteriors for both cases are shown in
Fig.~\ref{fig:mdc3result}. The estimated values of the Hubble 
constant are \Hubblemdcweights  \, \kmsMpc (assuming hosts are luminosity 
weighted) and \Hubblemdcnoweights  \, \kmsMpc (assuming equal weights).
By weighting the host galaxies with the correct function of their luminosities,
which happens to be known in this case,
the constraint on \ac{H0} improves --- the uncertainty narrows by a factor of 1.2,
compared to the case in 
which equal weights are assumed. Both results are consistent with the fiducial 
\ac{H0} value of 70 \kmsMpc.  In the limit of a far greater number of events, 
one might expect to see a bias emerge in the case in which the assumptions in 
the analysis do not match those with which the catalog was simulated. The 
luminosity weighting of host galaxies, by its very nature, increases the 
probability that the host galaxy is inside the galaxy catalog; assuming 
equal weighting gives disproportionate weight to the contribution that comes 
from beyond the galaxy catalog. However, for the 249 \ac{BNS} events considered 
here, the final posteriors are too broad to be able to detect any 
kind of bias.

\subsection{Comparison between the \acp{MDC} \label{sec:mdc comparison}}
\begin{figure*}
\includegraphics[width=0.48\linewidth]{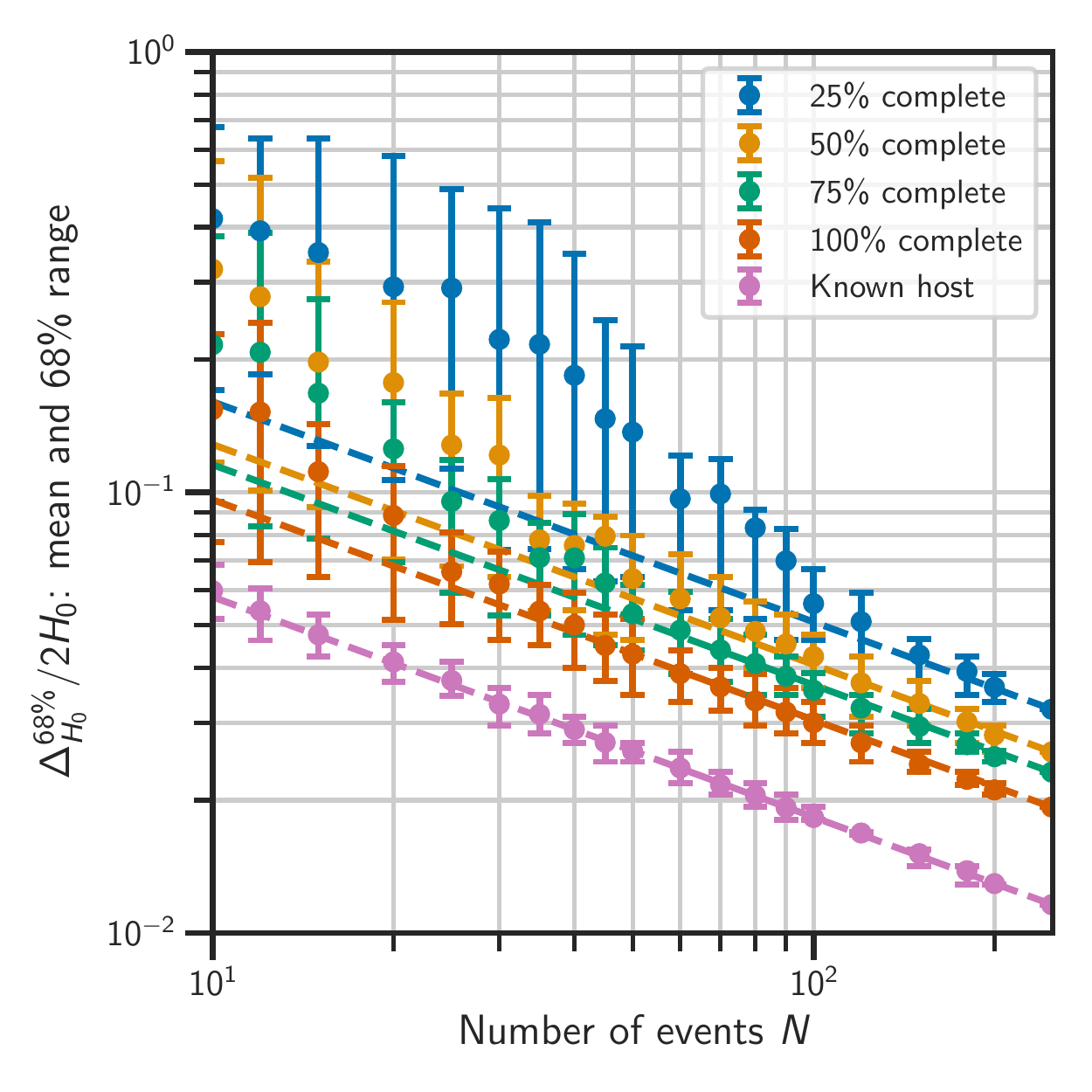}
\includegraphics[width=0.48\linewidth]{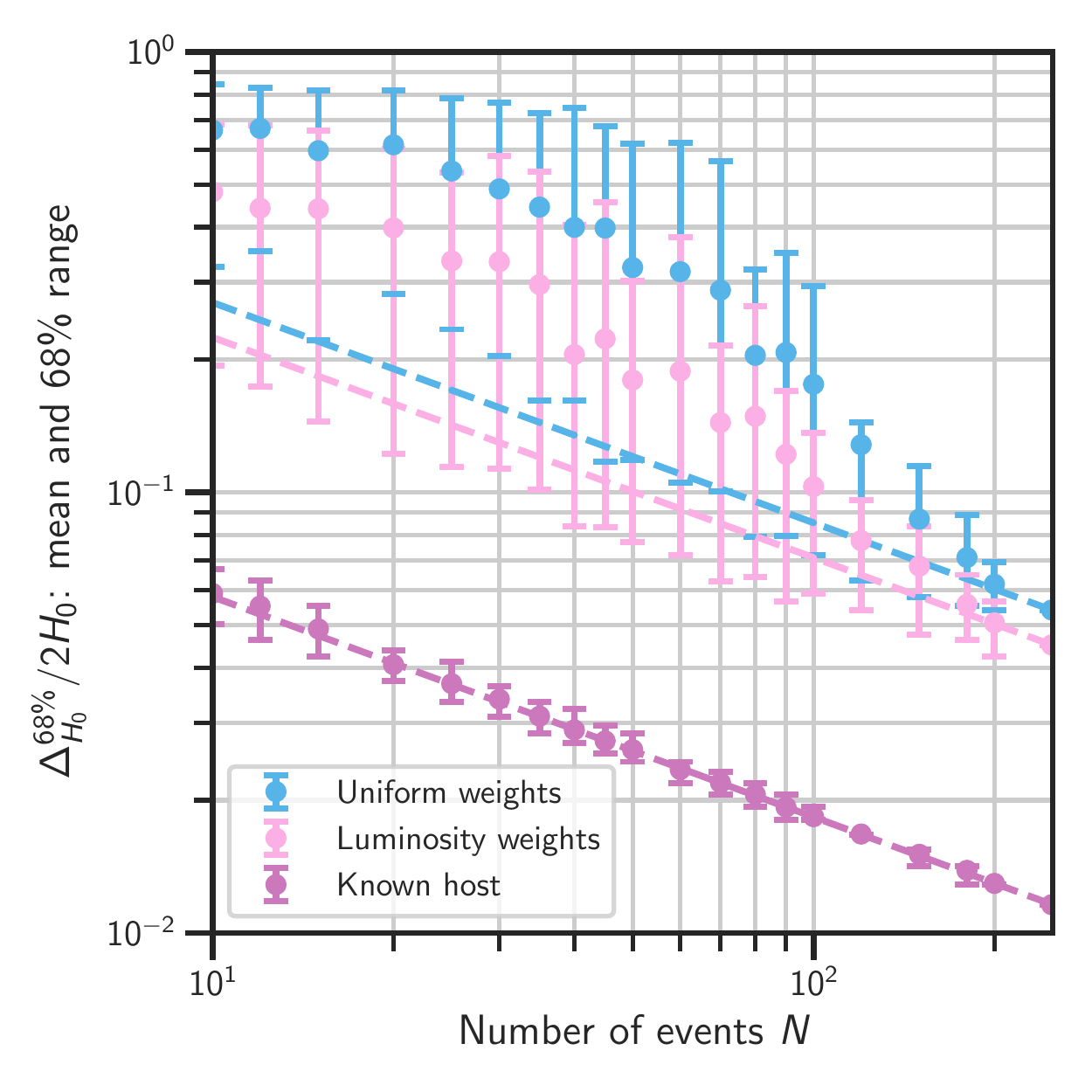}
\caption{\label{fig:mdcconvergence}
Fractional uncertainty in $H_0$  as a function of the number $N$ of the events for 
the combined $H_0$ posteriors. The fractional uncertainty in $H_0$ 
 is defined as the half-width of the 68.3\% highest probability density interval 
divided by $70$ \kmsMpc, 
and is shown as the plotted dots for all cases. The error bars contain 
68\% of the scatter arising from different realizations of the events.
(left) In purple, red, green, yellow and 
blue we show the associated host galaxy case (\ac{MDC}0), complete galaxy catalog 
(\ac{MDC}1) case, and the 75\%, 50\% and 25\% completeness cases; we find a 
fractional $H_0$ uncertainty of $\Hubblemdccountererror$\%, $\Hubblemdcerror$\%, 
$\Hubblemdccerror$\%, $\Hubblemdcberror$\% and $\Hubblemdcaerror$\% respectively for 
the combined $H_0$ posterior from 249 events. (right) convergence for \ac{MDC}3 
(event probability proportional to galaxy luminosity), analyzed with 
luminosity-weighted likelihood (pink) or equally-weighted likelihood (light 
blue).  We find fractional $H_0$ uncertainties of $\Hubblemdcweightserror$\% and $\Hubblemdcnoweightserror$\% respectively. \ac{MDC}0 (purple) is included for reference.
We plot the expected $1/ \sqrt{N}$ scaling behavior for large values of N for 
all cases with the dashed lines. This scaling behavior is met by all \acp{MDC} as 
the number of events reaches 249, but for the less informative, lower 
completeness \acp{MDC} the trend is slower to emerge. This is even more evident
in \ac{MDC}3, where the density of galaxies is 100 times greater, producing more 
potential hosts for each event. This is mitigated somewhat by the effect of 
luminosity-weighting the potential hosts (pink).
}
\end{figure*}

So far we have focused on individual event likelihoods and combined results for 
all 249 events. Our dataset also allows us to to assess the 
convergence for the combined Hubble posterior as we add events.
We calculate the intermediate combined posteriors as a function of the number of 
events, and show the resulting convergence in 
Fig.~\ref{fig:mdcconvergence}. 
We plot the fractional $H_0$ uncertainty (defined here as the half-width of the 
68.3\%
credible interval divided by $H_0$, $\Delta^{68.3\%}_{H0}/2H_0$), 
against the number of events we include in a randomly-selected group. The 
scatter between realizations of the group is indicated by the error bars, which 
encompass $68.3\%$ of their range.
There is a considerable variation between different realizations, for the 
incomplete catalogs. For example, of the 100 realizations we used, for 25\% 
completeness and 40 events, there are groups that give $\sim 10\%$ precision, 
but others that give $\sim 70\%$ precision.

With a sufficiently large number of events, we expect a $1/ \sqrt{N}$
scaling of the uncertainty with the number of events~\cite{Nissanke:2013fka,Chen:2017rfc}.
To check whether this behavior is indeed true, we fit the results for each \ac{MDC}
to the expected scaling, obtaining the coefficient of $1/\sqrt{N}$ by
maximizing its likelihood given the fractional uncertainties and their variances
from the different realizations. The coefficient of the scaling is automatically
dominated by the fractional uncertainties at large $N$ where the variances are small.
We show this scaling for each \ac{MDC} as a set of dashed lines in Fig.~\ref{fig:mdcconvergence}.

It can be seen that for each \ac{MDC}, the results converge to the expected 
$1/\sqrt{N}$ scaling. The number of events required before this behavior 
is reached is dependent on the amount of \ac{EM} information available 
on average for each event, in agreement with the results of~\cite{Chen:2017rfc}.
The direct counterpart case is always on the trend 
after $\mathcal{O}(10)$ events, and shows a $\sim18\%/\sqrt{N}$ convergence,
comparable to and consistent with the results in~\cite{Chen:2017rfc,Feeney:2018mkj}.
With the most complete galaxy 
catalogs, if the host galaxy is not directly identified it will take tens of 
events before this behavior is reached.  However, even the least complete 
catalog (25\%) appears to have reached this behavior by the time all 249 events 
are combined.  It should be noted that as the catalogs for \acp{MDC} 1 and 2 
were not simulated realistically, their low density relative to the density of 
the universe means that these numbers should not be taken as predictions of how 
fast $1/\sqrt{N}$ may be reached (except, perhaps, in the 
counterpart case, although one should bear in mind that even for that case, peculiar velocities and 
redshift uncertainties have been neglected). Even with a galaxy catalog which is
$25\%$ complete, \ac{MDC}2 gives a result which is only about a factor of $3$ times worse
than the counterpart case.

As the density of galaxies in \ac{MDC}3 was increased by 2 orders of magnitude 
over \acp{MDC} 1 and 2, the final posteriors cannot be directly compared between 
\acp{MDC}. However, by plotting the equivalent convergence figure for \ac{MDC}3 
(including the ``known host'' case as a reference, see 
Fig.~\ref{fig:mdcconvergence}), the impact of increasing the 
density of galaxies in the universe on the rate at which the posterior 
converges on the $1/\sqrt{N}$ behavior becomes clear. When there are more host galaxies, the 
results are overall less precise, and take longer to reach the $1/\sqrt{N}$
trend. As expected, using 
luminosity-weighting of potential host galaxies as an assumption in the 
analysis concentrates the probability to a smaller number of galaxies, leading 
to a more precise result.

\subsection{Limited Robustness Studies}

Our results are expected to be sensitive to the luminosity distribution parameters --- if one uses values of the Schechter function parameters $\alpha$ and $L^*$ in the analysis which are different from the ones used to simulate the galaxy catalogs, one would expect to end up with a bias in the results. With variations of these parameters within their current measurement uncertainties, we have however demonstrated that the resulting variation in the final result is small compared to the statistical uncertainties reached with the current set of \acp{MDC}. Furthermore we have also demonstrated that our results are robust against a small $\mathcal{O}(1)$ variation in the value of the telescope sensitivity threshold $m_\text{th}$.

\section{Conclusions and Outlook\label{sec:discussion}}
The $H_0$ measurement using \ac{GW} standard sirens has been demonstrated with recent 
events, including both the counterpart method for GW170817~\cite{GW170817:H0},
and the galaxy catalog method~\cite{DES,2018arXiv180705667F}. These 
approaches are combined in the analysis of both BNS and BBH events from the 
first two observing runs of the advanced detector network~\cite{O2H0paper},
using the method described in this paper.
Future measurements will rely on a combination of 
counterpart and catalog methods, as appropriate for each new detected event,
with catalog incompleteness playing an important role for the more distant, yet 
more common, \acp{BBH}.
This paper outlines a coherent approach that tackles both of these scenarios, 
including treatment of selection effects in both
\acp{GW} (due to the limited sensitivity of \ac{GW} detectors) 
and \ac{EM} (due to the flux-limitations of \ac{EM} observing channels).
We performed a series of \acp{MDC} to validate our method using up to 249 
observed events.
For each of the \acp{MDC} analyzed, the final 
posterior on \ac{H0} is found to be consistent with the value of 
$\ac{H0}=70\,\kmsMpc$ used to simulate the \ac{MDC} galaxy catalogs,
demonstrating that our method can produce sufficiently unbiased results for 
treating these numbers of events, in our simulations.

\ac{GW} selection effects are inherent in every version of the \ac{MDC} and were corrected for by the term $p(D_{\text{GW}}|H_0)$ in the denominator of Eq.~\eqref{Eq.xD}.
\ac{EM} selection effects are addressed in \acp{MDC} 2 and 3 by the 
out-of-catalog terms containing $\bar{G}$ in Eq.~\eqref{Eq:sum G}. In both these 
\acp{MDC}, in spite of having an apparent magnitude-limited galaxy catalog, we 
are able to accurately infer \ac{H0} without any bias.
\ac{MDC}2 further demonstrates our ability to account for missing host galaxies 
down to a level where only $25\%$ of events have hosts inside the catalog. Even 
in this case, we converge to the correct \ac{H0} value, to the level of 
precision which could be reached by 249 events.

\ac{MDC}3 demonstrates a clear tightening of the posterior distribution when we 
can assume that \ac{GW} events trace the galaxy luminosities, 
compared to the case in which we treat all galaxies as equally likely hosts.  
The ``uniform weights'' analysis of \ac{MDC}3 remains consistent with the 
simulated \ac{H0} value. Hence we are unable to conclude whether an incorrect 
assumption would lead to a biased result, as one might expect. We used only 249
events for our \acp{MDC}. With enough events of comparable nature the bias would
be detected. Future 
work will expand these studies to include a larger numbers of 
simulated \ac{GW} events, and will be able to discern smaller sources of systematic effects.

Although the galaxy-catalog standard-siren measurement of \ac{H0} is less precise 
than the counterpart measurement, it is still able to constrain $H_0$, but requires at least an order of magnitude more events in order to 
reach a comparable accuracy (in the most realistic case of \ac{MDC}3).
These \acp{MDC} have validated our method and implementation in 
simplified scenarios. However future work will be needed to improve on this in several directions, to test its applicability to \acp{BBH} (which are detectable out to much farther distances),
realistic cosmology, and real galaxy catalogs~\cite{O2H0paper,Chen:2017rfc}.

In both the counterpart and galaxy catalog cases, the lack of redshift uncertainties and peculiar 
velocities implies that the contributions from individual galaxies are a lot more 
precise than they would be in reality.
Moreover, the simulated catalogs in \acp{MDC} 
1 and 2 have a low density of galaxies compared to the universe, making them
more informative than real catalogs. 
\ac{MDC}3, with a galaxy density of 1 galaxy per 70 Mpc$^3$, comes closest to the actual density of 
galaxies in the local universe of $\sim1$ galaxy per 200 Mpc$^3$~\cite{GLADE}. In this scenario there is 
still a clear convergence towards the simulated \ac{H0} value. In comparison to 
actual catalogs such as GLADE~\cite{GLADE}, the apparent magnitude threshold of 
$14$ is very low, so we expect a real catalog-only analysis to fall somewhere 
between \acp{MDC} 2 and 3.
We caution the reader that with tens of events, the precision of results can vary by almost an 
order of magnitude depending on the particular realization of the detected 
population, before eventually converging to the expected $1/\sqrt{N}$ behaviour~\cite{Nissanke:2013fka,Chen:2017rfc}.
Analyzing more realistic catalogs will also require a sky-varying \ac{EM} selection 
function, as the magnitude threshold varies significantly on the sky according 
to the design of particular surveys.

The galaxy distribution in these simulated catalogs is uniform in 
comoving volume. 
Although it has not been studied here,
clustering of galaxies is expected to improve the constraint on \ac{H0} 
(see, {\em e.g.}~\cite{MacLeod:2007jd,Chen:2017rfc}), since even when the host is not in 
the catalog, it is likely that there will be observed galaxies nearby.

The Advanced LIGO - Virgo second observing run~\cite{O2:catalog} has confirmed 
that \ac{BBH} systems are detected at higher rates than \acp{BNS}. Since their 
greater mass allows them to be observed at much greater distances, where galaxy 
catalogs are incomplete, the catalog method including EM selection effects is 
particularly important. With the catalog of \ac{GW} 
events expected to expand at an increasing rate in future observing runs, our 
analysis will evolve to meet the challenges that come with it, and give us the fullest 
picture of cosmology as revealed by gravitational waves.

\begin{acknowledgments}
We thank members of the LIGO-Virgo Collaboration for valuable discussions pertaining to the writing of this paper,
and in particular Nicola Tamanini for a careful reading of the manuscript.
AG additionally thanks P.~Ajith, Walter Del Pozzo, Anuradha Samajdar, and Chris Van Den Broeck for discussion at various stages of the work.
RG, CM and JV are supported by the Science and Technology Research Council (grant No.ST/L000946/1).
IMH is supported by the NSF Graduate Research Fellowship Program under grant DGE-17247915. IMH also acknowledges support from NSF Grant No. PHY-1607585.
HQ is supported by Science and Technology Facilities Council (grant No.ST/T000147/1).
AS thanks Nikhef for its hospitality and support from the Amsterdam Excellence Scholarship (2016-2018).
HYC was supported by the Black Hole Initiative at Harvard University, through a grant from the John Templeton Foundation.
MF and DEH were supported by NSF grant PHY-1708081. They were also supported by the Kavli Institute for Cosmological Physics at the University of Chicago through an endowment from the Kavli Foundation.
AG is supported by the research programme of the Netherlands Organisation for Scientific Research (NWO). 
DEH gratefully acknowledges support from the Marion and Stuart Rice Award.
We are grateful for computational resources provided by the Leonard E Parker Center for Gravitation, Cosmology and Astrophysics at the University of Wisconsin-Milwaukee, and those provided by Cardiff University, and funded by an STFC grant supporting UK Involvement in the Operation of Advanced LIGO. This article has been assigned LIGO document number LIGO-P1900017.

\end{acknowledgments}

\bibliography{masterbib}

\appendix*
\begin{widetext}
\section{Detailed methodology\label{sec:methodappendix}}
\subsection{A note on luminosity weighting and redshift evolution \label{Sec: s term}}
The probability for a galaxy to host a \ac{GW} event is not uniform over all the galaxies present in the catalog.
Indeed, brighter galaxies are supposed to have an higher star-formation rate and hence have an higher probability to host a \ac{GW} event. Also galaxies at higher redshifts may be more likely to be hosts, as mergers are expected to be more frequent~\cite{Madau_2014}.
Our prior belief for a galaxy at redshift $z$, sky position $\Omega$ and absolute and relative magnitudes $M,m$,  to host a GW source $s$ can be expressed as
\begin{equation}
\label{Eq:expand_prior_start}
p(z,\Omega,M,m|s,H_0) =  p(m|z,\Omega,M,s,H_0)p(z,\Omega,M|s,H_0),
\end{equation}
where if we assume that $z,\Omega$ and $M$ are conditionally independent given $s,H_0$,
\begin{equation}
\label{Eq:expand_prior}
\begin{aligned}
p(z,\Omega,M,m|s,H_0) &=\delta(m - m(z,M,H_0))p(z|s)p(\Omega)p(M|s,H_0),
\\ &= \delta(m - m(z,M,H_0))\dfrac{p(s|z)p(z)}{p(s)}p(\Omega)\dfrac{p(s|M,H_0)p(M|H_0)}{p(s|H_0)}.
\end{aligned}
\end{equation}
In the last equation we used  the explicit relation between apparent magnitude and $z,M$ and $H_0$. 
The probability $p(z)$ is the prior distribution of galaxies in the universe, taken to be uniform in comoving volume-time, $p(\Omega)$ is the prior on galaxy sky location, assumed uniform over the celestial sphere, and $p(M|H_0)$ is the distribution of absolute magnitudes represented by the Schechter function. 
In the sections below we will show that the terms $p(s)$ and $p(s|H_0)$ cancel out with other terms, and so their exact form does not need to be considered.
$p(s|M,H_0)$ can take the form
\begin{equation}  \label{Eq:luminosityweighting}
\begin{aligned}
p(s|M,H_0) &\propto 
\begin{cases}
L(M(H_0)) & \text{if GW hosting probability is proportional to luminosity}\\
\text{constant} & \text{if GW hosting probability is independent of luminosity.}
\end{cases}
\end{aligned}
\end{equation}
We refer to the above equation as luminosity weighting.
The term $p(s|z)$ represents the probability for the merger rate to depend on the redshift,
\begin{equation}
\begin{aligned}
p(s|z) &\propto 
\begin{cases}
\text{function}(z) & \text{if rate evolves with redshift}\\
\text{constant} & \text{if rate is does not evolve with redshift.}
\end{cases}
\end{aligned}
\end{equation}
For the \acp{MDC} in this paper with $z\ll 1$, we assume a constant rate model but a more generic model with $p(s|z) \propto (1+z)^\lambda$ can be used with detections at higher redshifts. This was the case of ~\cite{O2H0paper}, for example, in which  a $p(s|z) \propto (1+z)^3$ was assumed.

\subsection{A detailed breakdown of the galaxy catalog case \label{Sec: Components}}

This section presents a more detailed look into the galaxy catalog method presented in section \ref{subsec:galcat method}.  The approach is summarized in Fig. \ref{fig:network}, a network diagram which shows how each of the parameters of this extended derivation fit together and their dependencies on each other.  The parameters which appear in this diagram, and in the following subsections, are defined in Table \ref{tab:params_ext}.

The subsections below provide derivations of the individual components of Eq. \ref{Eq:sum G}. Note that in the cases where the integration boundaries are not specified, they can be assumed to cover the full parameter space.

\begin{figure*}[t!]
\centering
\vspace{-1cm}
\includegraphics[width=\linewidth]{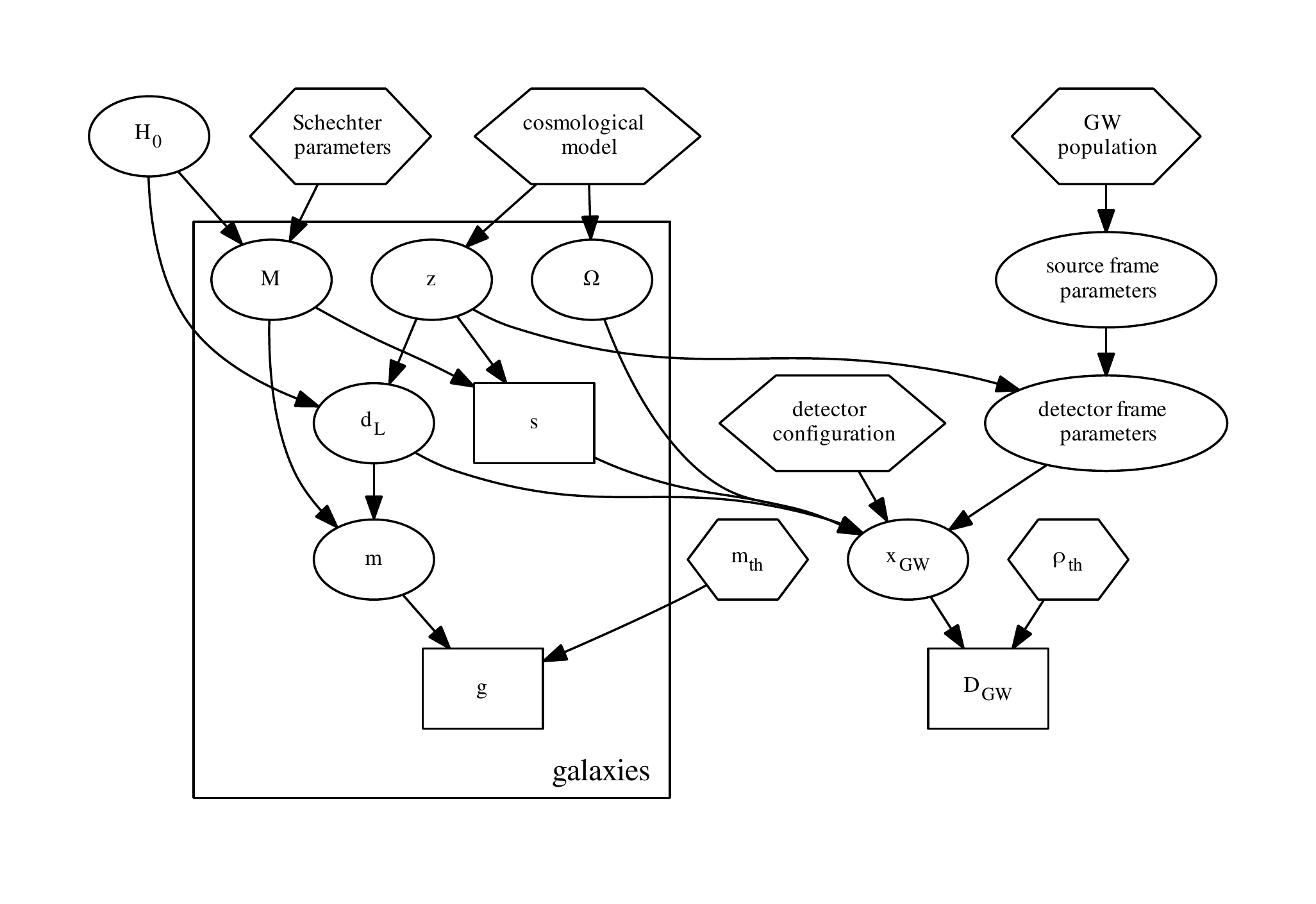}
 \vspace{-1.5cm}
\caption{\label{fig:network} A network diagram showing how the main parameters of the methodology interlink. Circular nodes denote ordinary parameters.  Hexagonal nodes denote assumed knowns.  Rectangular nodes denote binary flags. The arrows indicate the dependence of each parameter on the parameters which feed into them. The parameters grouped in the ``galaxies'' cluster are those which can be evaluated for every galaxy in the universe.}
\end{figure*}

{\renewcommand{\arraystretch}{1.3}
\begin{table}[b!]
\begin{ruledtabular}
\begin{tabular}{p{3.5cm}p{14.5cm}}
Parameter & Definition\\ \hline \hline
$H_0$ & The Hubble constant \\ \hline
$x_{\text{GW}}$ & The \ac{GW} data associated with some \ac{GW} source, $s$.  \\ \hline
$D_{\text{GW}}$ & Denotes that a \ac{GW} signal was detected, {\em i.e.}~that $x_{\text{GW}}$ passed some detection statistic threshold $\rho_\text{th}$. \\ \hline
$g$ & Denotes that a galaxy is ($G$), or is not ($\bar{G}$), contained within the galaxy catalog. \\  \hline
$s$ & Denotes that a \ac{GW} signal was emitted. \\ \hline
$M$ &  Absolute magnitude. \\ \hline
$z$  &  Redshift. \\ \hline
$\Omega$  &  Sky location (right ascension and declination). \\ \hline
$d_L$ &   Luminosity distance. \\ \hline
$m$  &  Apparent magnitude. \\ \hline
$m_{\text{th}}$ &   Apparent magnitude threshold of the galaxy catalog. \\ \hline
$\rho_{\text{th}}$ &  SNR threshold of the detector network. \\ \hline
Cosmological model & The cosmological model assumed for the analysis. Typically a Friedmann-Lema\^itre-Robertson-Walker universe. \\ \hline
Schechter parameters & The parameters which characterize the assumed absolute magnitude distribution of galaxies in the universe. \\ \hline
\ac{GW} population & The assumed underlying population of \ac{GW} sources. \\ \hline
Source frame parameters & Source frame parameters of a \ac{GW} source, {\em e.g.}~component masses, spins, inclination and polarization.  \\ \hline
Detector frame parameters & As above, but redshifted into the detector frame. \\ \hline
Detector configuration & The network set up, including which detectors are included in the search and their noise floors. \\
\end{tabular}
\end{ruledtabular}
\caption{\label{tab:params_ext} A summary of the parameters present in the network diagram, Fig. \ref{fig:network}.}
\end{table}
}

\subsubsection{Likelihood when host is in catalog: $p(x_{\text{GW}}|G,D_{\text{GW}},H_0)$}
\label{sec:app_simo}

The likelihood in the case where the host galaxy is inside the galaxy catalog, $p(x_{\text{GW}}|G,D_{\text{GW}},H_0)$, can be obtained from the marginalization over redshift, sky location, absolute magnitude and apparent magnitude. If we assume that the \ac{GW} data, $x_{\text{GW}}$, is independent of the galaxy catalog $G$, $m$ and $M$ we can write
\begin{equation}
\label{Eq:x_GDH0}
\begin{aligned}
p(x_{\text{GW}}|G,D_{\text{GW}},s,H_0) &= \dfrac{1}{p(D_{\text{GW}}|G,s,H_0)} \iiiint p(x_{\text{GW}}|z,\Omega,s,H_0) p(z,\Omega,M,m|G,s,H_0) dz d\Omega dM dm.
\end{aligned}
\end{equation}
The probability density function $p(z,\Omega,M,m|G,s,H_0)$ is taken as a sum of delta functions with specific $z$, $\Omega$ and $m$ corresponding to the location of each galaxy in the catalog. This can be further factorized as
\begin{equation}
\begin{aligned}
p(z,\Omega,M,m|G,s,H_0) &= \dfrac{p(s|z,\Omega,M,m,G,H_0)\delta(M - M(z,m,H_0))p(z,\Omega,m|G)}{p(s|G,H_0)},
\end{aligned}
\end{equation}
where we have assumed again a relation between the apparent magnitude, redshift, $H_0$ and absolute magnitude. This allows us to integrate over the absolute magnitude in Eq. \ref{Eq:x_GDH0} and obtain
\begin{equation}
\begin{aligned}
p(x_{\text{GW}}|G,D_{\text{GW}},s,H_0) &= \dfrac{1}{p(D_{\text{GW}}|G,s,H_0)p(s|G,H_0)}\iiint p(x_{\text{GW}}|z,\Omega,s,H_0) p(s|z,\Omega,M(z,m,H_0),m,G,H_0)p(z,\Omega,m|G) dz d\Omega dm.
\end{aligned}
\end{equation}
Remembering that $p(z,\Omega,m|G)$ represents the distribution of the galaxies in the catalog, we can replace the integral above with a sum over the galaxies.
\begin{equation}
\begin{aligned}
p(x_{\text{GW}}|G,D_{\text{GW}},s,H_0) &= \dfrac{1}{p(D_{\text{GW}}|G,s,H_0) p(s|G,H_0)} \dfrac{1}{N} \sum^N_{i=1} p(x_{\text{GW}}|z_i,\Omega_i,s,H_0) p(s|z_i)p(s|M(z_i,m_i,H_0)),
\end{aligned}
\end{equation}
where we have factorized $p(z_i|s)$ and $p(M(z_i,m_i,H_0)|s)$, together with the term $p(s|z,\Omega,M(z,m,H_0),m,G,H_0)$. Finally expanding the denominator $p(D_{\text{GW}}|G,s,H_0)$ in the same way, we can recover the likelihood for the ``in catalog'' part of the galaxy catalog method.
\begin{equation}
\begin{aligned}
p( x_{\text{GW}}|G, D_{\text{GW}}, s,H_0) &= \dfrac{\displaystyle \sum^N_{i=1} p(x_{\text{GW}}|z_i,\Omega_i,s,H_0) p(s|z_i)p(s|M(z_i,m_i,H_0))}{\displaystyle \sum^N_{i=1} p(D_{\text{GW}}|z_i,\Omega_i,s,H_0) p(s|z_i)p(s|M(z_i,m_i,H_0))}.
\end{aligned}
\end{equation}
Notably, in the case the galaxies in the catalogs are provided along with their redshift uncertainties $p(z_i)$, these can be implemented in the above equations as:
\begin{equation} \label{Eq:p(x|G,D,H0)}
\begin{aligned}
p(& x_{\text{GW}}|G, D_{\text{GW}}, s,H_0) = \dfrac{\displaystyle \sum^{N_\text{gal}}_{i=1} \int p(x_{\text{GW}}|z_i,\Omega_i,s,H_0) p(s|z_i) p(s|M(z_i,m_i,H_0)) p(z_i) dz_i}
{\displaystyle \sum^{N_\text{gal}}_{i=1} \int p(D_{\text{GW}}|z_i,\Omega_i,s,H_0) p(s|z_i) p(s|M(z_i,m_i,H_0)) p(z_i) dz_i}\,.
\end{aligned}
\end{equation}

\subsubsection{Probability the host galaxy is in the galaxy catalog: $p(G|D_{\text{GW}},H_0)$ and $p(\bar{G}|D_{\text{GW}},H_0)$}

The probability that the host galaxy is inside the galaxy catalog, given that a \ac{GW} signal was detected, can be expressed as
\begin{equation}
\label{Eq:G_DH0_start}
\begin{aligned}
p(G|D_{\text{GW}},s,H_0) &= \iiiint p(G|z,\Omega,M,m,D_{\text{GW}},s,H_0) p(z,\Omega,M,m|D_{\text{GW}},s,H_0) dz d\Omega dM dm,
\\ &= \iiiint \Theta[m_{\text{th}}-m] \dfrac{p(D_{\text{GW}}|z,\Omega,M,m,s,H_0) p(z,\Omega,M,m|s,H_0)}{p(D_{\text{GW}}|s,H_0)}  dz d\Omega dM dm ,
\\ &=  \dfrac{1}{p(D_{\text{GW}}|s,H_0)} \iiiint \Theta[m_{\text{th}}-m] p(D_{\text{GW}}|z,\Omega,s,H_0) p(z,\Omega,M,m|s,H_0) dz d\Omega dM dm.
\end{aligned}
\end{equation}
If we assume that the galaxy catalog is apparent magnitude-limited, such that only galaxies which are observed above some detection threshold are contained within it, we can approximate $p(G|z,\Omega,M,m,D_{\text{GW}},s,H_0)$ as a Heaviside step around  the detection threshold $m = m_{\text{th}}$.
\begin{equation}
\label{Eq:G_DH0_start_mid}
\begin{aligned}
p(G|D_{\text{GW}},s,H_0) &=\dfrac{1}{p(D_{\text{GW}}|s,H_0)} \iiiint \Theta[m_{\text{th}}-m] p(D_{\text{GW}}|z,\Omega,s,H_0) p(z,\Omega,M,m|s,H_0) dz d\Omega dM dm.
\end{aligned}
\end{equation}
We now expand $p(z,\Omega,M,m|s,H_0)$ as in Eq \ref{Eq:expand_prior} and we obtain
\begin{equation}
\label{Eq:G_DH0_mid}
\begin{aligned}
p(G|D_{\text{GW}},s,H_0) &= \dfrac{1}{p(s)p(s|H_0)} \dfrac{1}{p(D_{\text{GW}}|s,H_0)} \int^{z(M,m_{\text{th}},H_0)}_0 \! \! \! \! \! \! dz \int d\Omega \int dM p(D_{\text{GW}}|z,\Omega,s,H_0) p(s|z) p(z)p(\Omega)p(s|M,H_0)p(M|H_0).
\end{aligned}
\end{equation}
The term $p(D_{\text{GW}}|s,H_0)$ can be expanded in a similar way  and finally gives the probability for the host galaxy to be in the catalog.
\begin{equation}
\label{Eq:G_DH0_end}
\begin{aligned}
\\ p(G|D_{\text{GW}},s,H_0)&= \dfrac{\displaystyle \int^{z(M,m_{\text{th}},H_0)}_0 dz \int d\Omega \int dM p(D_{\text{GW}}|z,\Omega,s,H_0) p(s|z)p(z)p(\Omega)p(s|M,H_0)p(M|H_0)}{\displaystyle \iiint p(D_{\text{GW}}|z,\Omega,s,H_0) p(s|z)p(z)p(\Omega)p(s|M,H_0)p(M|H_0) dz d\Omega dM}.
\end{aligned}
\end{equation}
As the probabilities of being in the catalog and not in the catalog must be complementary, we have,
\begin{equation}
\begin{aligned}
p(\bar{G}|D_{\text{GW}},s,H_0) &= 1 - p(G|D_{\text{GW}},s,H_0).
\end{aligned}
\end{equation}

\subsubsection{Likelihood when host is not in catalog: $p(x_{\text{GW}}|\bar{G},D_{\text{GW}},H_0)$}
We follow an approach similar to the one presented in Appendix \ref{sec:app_simo}. We expand
\begin{equation}
\label{Eq:px_H0GbarD}
\begin{aligned}
p(x_{\text{GW}}|\bar{G},D_{\text{GW}},s,H_0) &= \dfrac{1}{p(D_{\text{GW}}|\bar{G},s,H_0)} \iiiint p(x_{\text{GW}}|z,\Omega,s,H_0) \dfrac{p(\bar{G}|z,\Omega,M,m,s,H_0)p(z,\Omega,M,m|s,H_0)}{p(\bar{G}|s,H_0)} dz d\Omega dM dm,
\end{aligned}
\end{equation}
The prior term, $p(z,\Omega,M,m|s,H_0)$ can now be expanded as it was in Eq \ref{Eq:expand_prior}. Substituting this in, and utilizing a Heaviside step function to represent the galaxy catalog's apparent magnitude threshold for $p(\bar{G}|z,\Omega,M,m,s,H_0)$,
\begin{equation}
\begin{aligned}
p(x_{\text{GW}}|\bar{G},s,H_0) &=\dfrac{1}{p(s)p(s|H_0)} \dfrac{1}{p(\bar{G}|s,H_0)}\int^\infty_{z(H_0,m_{\text{th}},M)} \! \! \! \! \! dz \int d\Omega \int dM p(x_{\text{GW}}|z,\Omega,s,H_0) p(s|z)p(z)p(\Omega)p(s|M,H_0)p(M|H_0).
\end{aligned}
\end{equation}
Expanding the denominator, $p(D_{\text{GW}}|\bar{G},s,H_0)$, in the same way gives an equivalent term,
\begin{equation}
\begin{aligned}
p(D_{\text{GW}}|\bar{G},s,H_0) &=\dfrac{1}{p(s)p(s|H_0)} \dfrac{1}{p(\bar{G}|s,H_0)}\int^\infty_{z(H_0,m_{\text{th}},M)} \! \! \! \! \! dz \int d\Omega \int dM p(D_{\text{GW}}|z,\Omega,s,H_0)  p(s|z)p(z)p(\Omega)p(s|M,H_0)p(M|H_0).
\end{aligned}
\end{equation}
And substituting this back into Eq \ref{Eq:px_H0GbarD} finally gives,
\begin{equation}
\label{Eq:p(x|barG,D,H0)}
\begin{aligned}
p(x_{\text{GW}}|\bar{G},D_{\text{GW}},s,H_0) &= \dfrac{\displaystyle \int^\infty_{z(M,m_{\text{th}},H_0)}dz \int d\Omega \int dM p(x_{\text{GW}}|z,\Omega,s,H_0) p(s|z)p(z)p(\Omega)p(s|M,H_0)p(M|H_0)}{\displaystyle \int^\infty_{z(M,m_{\text{th}},H_0)}dz \int d\Omega \int dM p(D_{\text{GW}}|z,\Omega,s,H_0) p(s|z)p(z)p(\Omega)p(s|M,H_0)p(M|H_0)}.
\end{aligned}
\end{equation}

\subsection{The catalog patch case}
While in general the galaxy catalog method derived in \ref{Sec: Components} was for use with a galaxy catalog which covers the entire sky, a small modification allows the use of catalogs which only cover a patch of sky, as long as the patch can be specified using limits in right ascension and declination.  If we represent the sky area covered by the catalog as $\Omega_{\text{cat}}$, and the area outside the catalog as $\Omega_{\text{rest}}$, such that $\Omega_{\text{cat}}+\Omega_{\text{rest}}$ covers the whole sky, this can be written as follows:
\begin{equation}
\begin{aligned}
p(x_{\text{GW}}|D_{\text{GW}},H_0) &= \int p(x_{\text{GW}}|\Omega,D_{\text{GW}},H_0)p(\Omega) d\Omega,
\\&=  \int^{\Omega_{\text{cat}}} p(x_{\text{GW}}|\Omega,D_{\text{GW}},H_0)p(\Omega) d\Omega + \int^{\Omega_{\text{rest}}}p(x_{\text{GW}}|\Omega,D_{\text{GW}},H_0)p(\Omega) d\Omega.
\end{aligned} 
\end{equation}
The first term is equivalent to the regular galaxy catalog case, but with limits on the integral over $\Omega$, while the second term has no $G$ and $\bar{G}$ terms, and covers the rest of the sky from redshift 0 to $\infty$.

\subsection{Direct and pencil beam counterpart cases \label{Ap:counterpart}}

The ``direct'' method assumes that the counterpart has been unambiguously linked to the host galaxy of the \ac{GW} event, such that the redshift and sky location of that galaxy can be taken to be that of the \ac{GW} event with certainty, see Eq. \ref{Eq:counterpart}.  Instead the numerator is calculated by evaluating the \ac{GW} likelihood at the delta-function location of the counterpart in $z$ and $\Omega$, and the term in the denominator is evaluated as:
\begin{equation}
p(D_{\text{GW}}|H_0) = \iiint p(D_{\text{GW}}|z,\Omega,H_0) p(z)p(\Omega)p(M|H_0) dz d\Omega dM,
\end{equation} 
for priors $p(z)$ and $p(\Omega)$ (note that this is independent of galaxy catalog data).

The ``pencil-beam'' method makes the assumption that while the sky location of the galaxy associated with the counterpart is that of the GW event, we may not make a direct association to a known galaxy but to a set of potential candidate hosts. We can use the EM constrained sky localization and therefore return to the question of whether the host is within or beyond the galaxy catalog.  In this case, the likelihood takes the same form as in the galaxy catalog case, but evaluated along the line of sight of the candidate counterparts.

\subsection{GW selection effects \label{Ap:GWselection}}
Eq. \ref{Eq:D_H0} in section \ref{Sec: Overview} can be written as:
\begin{equation}
\begin{aligned}
p(D_{\text{GW}}|H_0) = \int p(D_{\text{GW}}|x_{\text{GW}},H_0)p(x_{\text{GW}}|H_0)dx_{\text{GW}}.
\end{aligned}
\end{equation}
where $p(D_{\text{GW}}|x_{\text{GW}},H_0)$ is a binary quantity which is 1 if the \ac{SNR} of $x_{\text{GW}}$ passes $\rho_{th}$, and 0 otherwise.

Looking at the individual components of Eq. \ref{Eq:sum G} in their expanded forms ({\em e.g.}~Eq. \ref{Eq:p(x|G,D,H0)}, \ref{Eq:G_DH0_end} and \ref{Eq:p(x|barG,D,H0)}), $p(D_{\text{GW}}|H_0)$ only appears in an expanded form, where it is additionally conditioned on $z$ and $\Omega$.
Calculating $p(D_{\text{GW}}|z,\Omega,H_0)$ requires integrating over all realizations of \ac{GW} events (detected and not), for a range of $z$, $\Omega$ and $H_0$ values, and applying a detection threshold ($\rho_{th}$) which all events must pass in order to be deemed detected.

Practically, Monte-Carlo integration can be used:
\begin{equation}
\begin{aligned}
p(D_{\text{GW}}|z,\Omega,H_0) = \dfrac{1}{N_{\text{samples}}} \sum^{N_{\text{samples}}}_{i=1} p({D_{\text{GW}}}_i|{x_{\text{GW}}}_i,z,\Omega,H_0).
\end{aligned}
\end{equation}
where ${x_{\text{GW}}}_i$ corresponds to an event, the parameters of which have been randomly drawn from the prior distributions of parameters which affect an event's detectability (mass, inclination, polarization, and sky location) and the event's $\rho_i$ is calculated for specific values of $z$ and $H_0$.
\begin{equation}
\begin{aligned}
p({D_{\text{GW}}}_i|{x_{\text{GW}}}_i,z,\Omega,H_0)=
\begin{cases}
1, & \text{if} \ \rho>\rho_{th}\\
0, & \text{otherwise.}
\end{cases}
\end{aligned}
\end{equation}
which gives a smooth function for $p(D_{\text{GW}}|z,\Omega,H_0)$, which drops from 1 to 0 over a range of $z$, $\Omega$ and $H_0$ values.

For the \ac{MDC}, we use a $\rho_{th}$ of 8 in each detector (assuming every event was detected by two detectors) and the 2016 PSD from \cite[Fig. 1]{2014ApJ...795..105S}, and evaluate $p({D_{\text{GW}}}_i|{x_{\text{GW}}}_i,z,\Omega,H_0)$ for 5000 samples, such that the integral converges. For this analysis, we assume that the probability of detection is averaged over the course of the entire simulated observation period, such that the dependence of $D_{\text{GW}}$ on $\Omega$ is smeared out over the course of many days.  We approximate this to mean that $p(D_{\text{GW}}|z,H_0)$ is uniform over the sky (ignoring the mild declination dependence which would remain after the rotation of the Earth is taken into account).  Fig.~\ref{fig:pdet} shows how the probability of detection behaves as a function of $z$ for different values of $H_0$.

\begin{figure*}
\includegraphics[width=0.75 \linewidth]{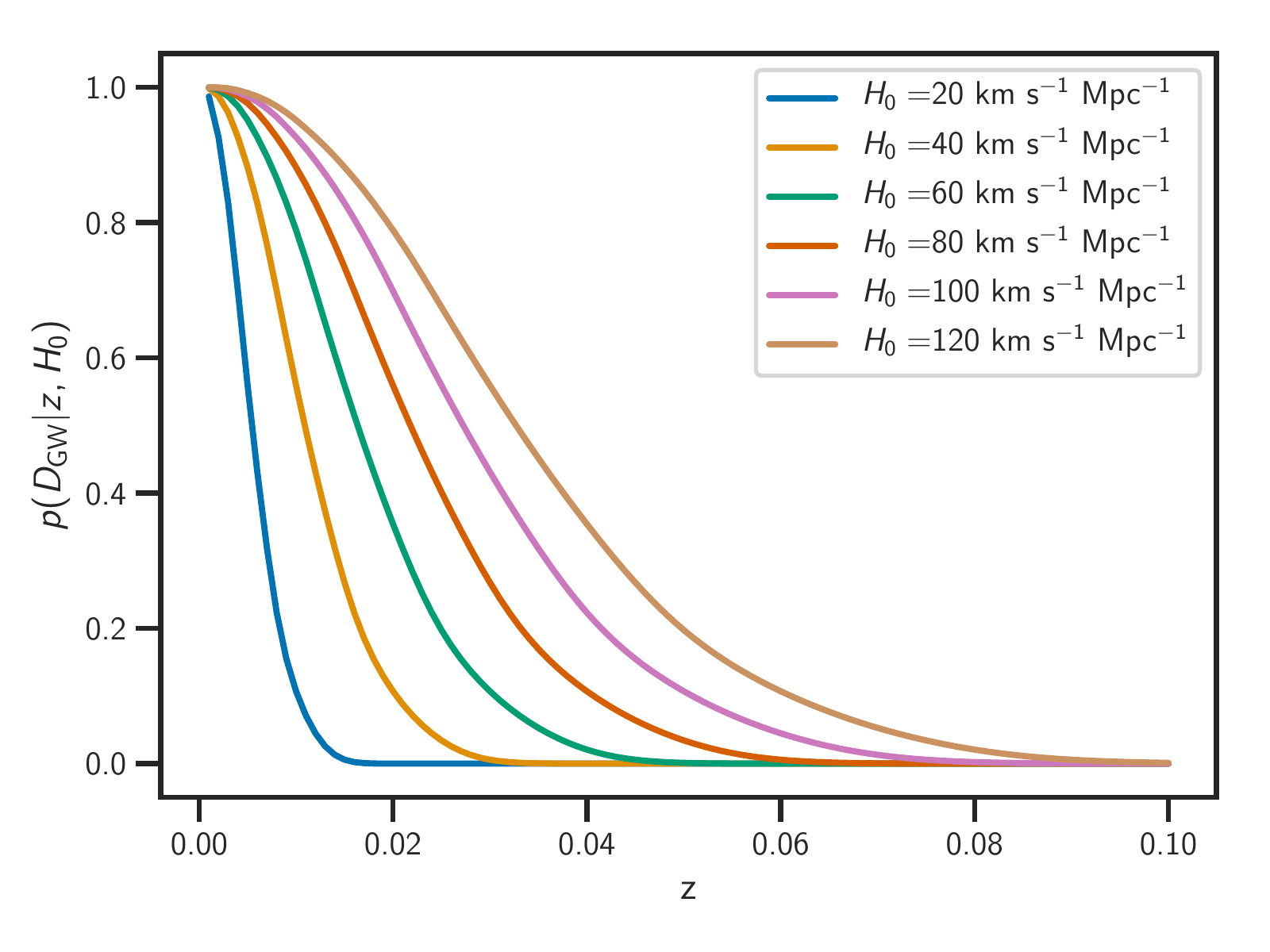}
\caption{Probability of detection, $p(D_{\text{GW}}|z,H_0)$, as a function of $z$ for different values of $H_0$.  We assume a 2-detector network at \ac{O2}-like sensitivity, for a population of binary neutron stars.}
\label{fig:pdet}
\end{figure*}

\subsection{Prior mass distribution}

An event's detectability is dependent on its observed (redshifted) detector-frame mass, $M_z$, but priors on the mass refer to their source-frame mass.  When calculating $p(D|H_0)$ the masses are drawn from the priors on source mass, $p(M_1,M_2)$ and then converted to observed masses through the equation:
\begin{equation}
\begin{aligned}
M_z = (1+z)M.
\end{aligned}
\end{equation}
However, when we use \ac{GW} data in the form of posterior samples, the prior 
used to generate those is uniform on the redshifted mass, $M_z$ 
\cite{LALInference}.  Due to the way the \ac{MDC} \ac{GW} data was generated, 
with masses chosen on the detector-frame, rather than the source-frame, this was 
not something which had to be considered. With real \ac{GW} data, as the 
redshift is linked directly to $H_0$, it is necessary to take into account the 
redshifting of the masses explicitly.

In general, when calculating $p(D|H_0)$ for BBHs, the primary mass $M_1$ is 
drawn from a power-law with slope $\alpha$, between limits $[a,b] \ M_{\odot}$.  
The secondary mass, $M_2$ is drawn from a uniform distribution between 
$aM_{\odot}$ and $M_1$ \cite{O2:Rates}, to give (for $\alpha \neq -1$):
\begin{equation}
\begin{aligned}
p(M_1,M_2) = \dfrac{(\alpha+1)M_{1}^\alpha}{bM_{\odot}^{(\alpha+1)}-aM_{\odot}^{(\alpha+1)}} \dfrac{1}{M_{1}-aM_{\odot}}.
\end{aligned}
\end{equation}

This is related to the redshifted mass by the Jacobian:
\begin{equation}
\begin{aligned}
p(M_{1,z},M_{2,z}) &= p(M_{1},M_{2}) \bigg\vert \dfrac{\partial(M_{1},M_{2})}{\partial(M_{1,z},M_{2,z})} \bigg\vert,
\\ &= p(M_{1},M_{2}) \bigg\vert \dfrac{1}{(1+z)^2} \bigg\vert.
\end{aligned}
\end{equation}
Substituting in our expression for $p(M_{1},M_{2})$:
\begin{equation}
\begin{aligned}
p(M_{1,z},M_{1,z}) &= \dfrac{(\alpha+1)M_{1}^\alpha}{bM_{\odot}^{(\alpha+1)}-aM_{\odot}^{(\alpha+1)}} \dfrac{1}{M_{1}-aM_{\odot}} \dfrac{1}{(1+z)^2},
\\ &= \dfrac{(1+z)^2(\alpha+1)M_{1,z}^\alpha}{bM_{\odot, z}^{(\alpha+1)}-aM_{\odot, z}^{(\alpha+1)}} \dfrac{1}{M_{1,z}-aM_{\odot, z}} \dfrac{1}{(1+z)^2},
\\ &= \dfrac{(\alpha+1)M_{1,z}^\alpha}{bM_{\odot, z}^{(\alpha+1)}-aM_{\odot, z}^{(\alpha+1)}} \dfrac{1}{M_{1,z}-aM_{\odot, z}}.
\end{aligned}
\end{equation}
The factor of $(1+z)^2$ cancels in the numerator and denominator.  As all redshift (and hence $H_0$) dependence has been removed, no correction is required for the differing priors.  For the case in which $\alpha=-1$, it can be shown that all redshift dependence falls out as well, meaning that as long as the prior mass distribution takes the form of a power-law, no prior correction is required. This will not be the case for all mass distributions.

\end{widetext}

\end{document}